\documentclass{aa}
\usepackage{xcolor}

\usepackage{graphicx}
\usepackage{txfonts}
\usepackage{natbib}
\usepackage{float}
\RequirePackage{etex}
\begin{document}

   \title{Direct estimation of the lunar-solar precession with geodetic very long baseline interferometry}

   \author{O. Titov$^{1,2}$, A. Osetrova$^3$
                    }

   \institute{$^1$Shanghai Astronomical Observatory,
              80 Nandan road, Shanghai, 200030, China\\
              \email{oleg.titov@shao.ac.cn} \\
              $^2$ Phase$\&$Rate, 2600, ACT, Australia\\
              $^3$  Institute of Applied Astronomy of the Russian Academy of Sciences, Kutuzov embankment, 10, Saint Petersburg, Russia 
             }

   \date{Received June .., 2026}

\abstract
{
\emph{Context}. Geodetic very long baseline interferometry (VLBI) is capable of measuring the Earth's instantaneous rotation pole (IRP) components (hereafter, $X_{IRP}$ and $Y_{IRP}$)  with sufficient accuracy using the VLBI delay rate. The least-squares fit of the $X_{IRP}$ and $Y_{IRP}$ polynomial expansion coefficients produces the precession model parameters independently of planetary motion theories. 

\emph{Aims}. Our aim was to estimate the precession parameters using VLBI delay rate observations as well as the nutation coefficients in order to produce a full transformation matrix from the IRP to the Geocentric Celestial Reference System (GCRS).  Then, the goal was to incorporate the free core nutation (FCN) effect from the VLBI group delay to form a complete and simple analytical model.

\emph{Results}.  
We obtained an empirical model of the transformation from International Terrestrial Reference System (ITRS) to Geocentric Celestial Reference System (GCRS) based on observational results. Similar to the IAU2006 theory, the new model produces time series of $X_{IRP}$ and $Y_{IRP}$ angles, but these two angles now point out the direction of the IRP rather than the celestial intermediate pole (CIP). This new model  includes a quadratic (or cubic) trend for precession and a set of periodic signals for nutation, correspondingly. We propose a separate model for the FCN nutation to replace the standard tabulation approach.

\emph{Conclusion}.
The IRP could be defined empirically with the $X_{IRP}$ and $Y_{IRP}$ values obtained directly from VLBI observations. A simple analytical model defines the IRP location for any epoch during the interval of observations. Therefore, the non-rotation origin concept is not required for the transformation matrix construction.
The estimate of the lunar-solar precessional parameter based on analysis of VLBI delay rate is $\psi_{A}= 5038.500\pm 0.019$~$''$/cy (referred to the observational epoch 2000.0), which is close to the adopted IAU2006 precessional parameter value $\psi_{A}= 5038.4815 $~$''$/cy.
}
 
\keywords{geodetic VLBI -- Earth rotation -- precession -- free core nutation}

\maketitle 

\section{Introduction}

Geodetic very long baseline interferometry (VLBI) measures the difference in arrival times of radio waves --- ‘group delay’ --- at two radio telescopes receiving signals from distant sources and a relative fringe frequency (also known as the 'delay rate') that compensates for the geocentric motion of each radio telescope due to the Earth's rotation. This measurement achieves remarkable precision, with accuracies of 20-40 picoseconds (ps) for the group delay and 1-100 frad for the fringe frequency \citep{Schuh12}.

While the group delay is widely used for estimating a large set of geodetic and astrometric parameters, the fringe frequency is almost entirely ignored due to the precision limitations. However, the fringe frequency could be used as a direct measure of the instantaneous Earth rotation vector, which is not accessible with the group delay.

Transformation of the baseline vector from the terrestrial to celestial reference frame requires an introduction of the so-called celestial intermediate pole (CIP). The group delay measured with the geodetic VLBI outputs the polar motion with respect to the CIP in the International Terrestrial Reference Frame (ITRF) using the three parameters as two polar motion components (known as x- and y-pole motion components) and the Universal Time (UT1), which could be compared to the Universal Coordinated Time (UTC). The difference between the UT1 reflecting the irregularities of the Earth rotation and the more homogeneous atomic-based UTC is monitored by the International Earth Rotation Service (IERS). The ITRF pole is linked to the Geocentric Celestial Reference System (GCRS) pole with the precession-nutation model as well as polar motion. Therefore, the full set of five parameters is estimated from a typical 24-hour geodetic VLBI experiment.

The precession and nutation transformation is a standard way to model the motion of the Earth's rotational axis in the inertial space and calculate the apparent equatorial positions of the celestial bodies on any date.  The traditional method used over centuries is valid for the slow motion of the reference point (equinox), which depends on the motion of the Earth's rotation and the slower motion of the ecliptic due to the so-called precession of the ecliptic. The modern, officially adopted alternative is based on the concept of the non-rotating origin, which defines the motion of the CIP in a certain time interval. The latter method uses only three parameters: $X$, $Y$,  which define the orientation of the CIP, and $s$, a celestial intermediate origin (CIO) locator, which is a function of $X, Y$ and can be calculated numerically. 

Historically, precession was discovered by the Greek astronomer Hipparchus in the second century BCE. He noted that positions of stars slightly change with time when comparing two stellar catalogues separated by two hundred years. Hipparchus estimated a precession of as much as $50.3''$  per year (\cite{Dambis_2000}, \cite{Vavin_2022}). More recently, \cite{Fricke_1971} and \cite{Lieske_1985} presented detailed historical analysis of numerous attempts to improve the precessional constant since the early 19th century. All the results were based on the observations of the stellar proper motion.
\cite{Lieske_1985} published the final value based on the stellar astrometry as $\psi_{A} = 5038.7784''/\text{cy}$ for the basic epoch J2000 but with linear trend of $-1.07259''/\text{cy}^2$. Therefore, the constant precession rate referring to the epoch J2000.0 is not really constant; it actually changes slowly over time. 

The modern precessional constant adopted by the International Astronomical Union (IAU) 2006 General Assembly \cite{iers10} is equal to $\psi_{A} = 5038.481507''/\text{cy}$, and it relies on the theoretical ephemerides of the Solar System planets, Earth's dynamical structure, the precession rate from the IAU2000 (MHB) model, and the CIP concept. The observational verification of the theoretical results was stated to be insufficient due to a lack of observational data by 2003 and some complications of the data analysis (e.g. \cite{Mathews2002}, \cite{Capitaine_2003}, \cite{capitaine_2005}). Since the theoretical precession and nutation values are applied for the reduction of VLBI data, the minor corrections to the a priori values are fitted using the least-squares method and referred to as corrections to the nutation model, whereas the precession model is not evaluated (\citet{Krasna_2013, Malkin_2013, Belda2017}). New development of theory based on the Earth’s dynamical flattening $J_{2}$ temporal dependence obtained with satellite laser ranging helps reduce the long-term variability of the nutation time series, but it leaves a potential for further improvement (\cite{Liu_2017}, \cite{Escapa_2017}, \cite{Baenas_2019}, \cite{Liu_2025}). The new approach developed in this paper suggests an opportunity for estimation of the precessional parameters free of
the theoretical assumption and most of the observational limitations effective at the dawn of the 21st century.

The plan of this paper is as follows: 1) to show the transition from the VLBI delay rate observations to the {$X_{IRP}$ and $Y_{IRP}$ time series, 2) to calculate the $X_{IRP}$ and $Y_{IRP}$ time series from the VLBI delay rate observations over a long time range, 3) to estimate the precession parameters and main nutation harmonics with the newly calculated {$X_{IRP}$ and $Y_{IRP}$ time series, 4) to elaborate a simple analytical model for free core nutation (FCN) calculated with VLBI group delay observations, and 5) to combine all parameters together within a simple method of transformation between ITRS and GCRS.

\section{Link of the VLBI delay rate and precession}

The positions of the CIP in the $X$ and $Y$ directions are predicted by the IAU2000A/IAU2006 nutation and precession models (\citet{Mathews2002}; \citet{Capitaine2003};  \citet{Capitaine2008}; \citet{Capitaine_2009}). The standard approach focuses on analysing group delay and applying corrections to the rotation angle values ($dX$, $dY$) in order to calculate minor adjustments 
to precession and nutation amplitudes (e.g. \citet{Capitaine_2009}; \citet{iers10}; \citet{Belda2017}).  The least-squares method is then employed to solve 
the system of equations after incorporating all relativistic corrections. Typically, this method 
requires a priori information about the estimated Earth orientation parameters (EOP). 
While it is often asserted that geodetic techniques (VLBI included) cannot measure the $X$ and $Y$ angles directly, \cite{Titov_2025} has shown that the fringe frequency measured with geodetic VLBI, as described in this section below, can be used to provide an accurate direction of the  Earth's angular rotation vector, which defines the position of the instantaneous rotation pole (IRP).

As is shown below, the difference between the IRP and CIP is subtle and mostly conceptual. While the CIP relies on the theoretical condition, the IRP is estimated with some formal uncertainty from VLBI observations. The practical consequence is in the calculation of the angle, which is known as the CIO locator ($s$) in the non-rotating origin concept, and the IRP locator ($S$) in the approach proposed in this paper.

\cite{Titov_2025} has shown that the three components ($\omega_{x}, \omega_{y}, \omega_{z}$) of the Earth's instantaneous rotation vector $\Omega$ could be estimated and presented as follows

\begin{equation}
\begin{aligned}
&\omega_{x} = \Omega \sin X_{IRP} \\
&\omega_{y} = \Omega \cos X_{IRP} \sin Y_{IRP}  \\
&\omega_{z} = \Omega \cos X_{IRP} \cos Y_{IRP} \\
&\Omega = \sqrt{\omega^{2}_{x} + \omega^{2}_{y} + \omega^{2}_{z}} ,
\end{aligned}
	\label{Eq1}   
\end{equation}
where $X_{IRP}$ and $Y_{IRP}$ are coordinates of the IRP. One can then calculate the IRP components $X_{IRP}$ and $Y_{IRP}$ from Eq.~(\ref{Eq1}):

\begin{equation}
\begin{aligned}
&X_{IRP} =  \arctan\frac{\omega_{x}}{\sqrt{\omega^2_{y}+\omega^2_{z}}}\approx \frac{\omega_{x}}{\sqrt{\omega^2_{y}+\omega^2_{z}}} \\
&Y_{IRP} = \arctan \frac{\omega_{y}} {\omega_{z}}  \approx \frac{\omega_{y}} {\omega_{z}} \\
&Z_{IRP} = \sqrt{1 - X_{IRP}^2 - Y_{IRP}^2}. \\
\end{aligned}
	\label{Eq2}  
\end{equation}

Then, based on Eq.~(\ref{Eq2}), formal errors of $X_{IRP}$ and $Y_{IRP}$ are given by

\begin{equation}
\begin{aligned}
&\sigma_{X_{IRP}} = \frac{ \sqrt{ (\omega_y^2 + \omega_z^2)\sigma_{\omega_x}^2 + \frac{\omega_x^2 \omega_y^2}{\omega_y^2 + \omega_z^2}\sigma_{\omega_y}^2 + \frac{\omega_x^2 \omega_z^2}{\omega_y^2 + \omega_z^2}\sigma_{\omega_z}^2 }}{\Omega^2}  \\ 
&\sigma_{Y_{IRP}} = \frac{\sqrt{\omega_z^2 \sigma_{\omega_y}^2 + \omega_y^2 \sigma_{\omega_z}^2}}{\omega_y^2 + \omega_z^2}. 	 
\end{aligned}
\label{Eq3}
\end{equation}
It is important to note that the VLBI group delays (used in the standard method) are affected by hydrogen maser phase fluctuation offsets and the local troposphere wet component instability. 
This necessitates the development of complex algorithms to mitigate the effects of phase variations when using standard group delay for data analysis. In contrast, the fringe frequency in Eq.~(\ref{Eq2}) is not affected by such phase fluctuations. 
Therefore, a simple least-squares method is effective, and it can yield daily estimates of the three components of the vector $\Omega$  for each 24-hour VLBI experiment.

\section{Results}

\subsection{Solution setup}

We processed the VLBI data used in the operational service from April 1983 to April 2026 
(NEOS-A in 1993-2001 and IVS-R1, R4 in 2002-2026) with the OCCAM 
software platform (version 6.3) \citep{Titov2004}
to estimate the three components of the vector $\Omega$. All standard reductions of geodetic VLBI data adhere to the IERS conventions of 2010 \citep{iers10}. Station positions and 
velocities were fixed based on ITRF2020 values \citep{Altamimi2023}, with a priori EOP sourced from IERS ~C04 data and the IAU 2000A nutation model. Radio source positions were anchored by the ICRF3 catalogue coordinates \citep{Charlot_2020}. 
We used the recent model for the ocean loading effect developed by M.S. Bos and H.-G. Scherneck from the Onsala Space Observatory (OSO)\footnote{\url{http://holt.oso.chalmers.se/loading/index-aside-2404271219.html}}.
A priori zenith delays were determined from local troposphere parameter values and subsequently mapped to the observation elevation using the Vienna mapping function (VMF1) \citep{Boehm2006}.

\subsection{Estimation of the long-term trends from the VLBI delay rate}

The components $X_{IRP}$, $Y_{IRP}$ obtained over the 43-year period of observations using Eqs~(\ref{Eq1}) - (\ref{Eq3}) are shown in Figure~\ref{fig1}. The angular velocity $\Omega$ and component Z were also calculated, and the details may be found in the previous paper \cite{Titov_2025}. The formal errors $\sigma_{X_{IRP}}$  and $\sigma_{Y_{IRP}}$ are so small that they cannot be noticed in Figure~\ref{fig1}; we present them separately in Figure~\ref{fig1a}.

\begin{table}[!h]
\caption{Estimates of the precession model.} 
\label{polynom}
\begin{tabular}{c|c|c|c} 
\hline
\noalign{\smallskip}
i & $a_{i}$ & $a_{i}$   & units \\
& & IERS 2010 & \\
\noalign{\smallskip}
\hline 
\noalign{\smallskip}
0 & $ -0.0478 \pm 0.0007$  &  $ -0.016617 $ & $''$  \\     
1 &    $2004.182 \pm 0.077 $ & $ 2004.191898 $ & $''/cy$  \\
2 &  $-0.590 \pm 0.041$  &  $ -0.4297829 $ & $''/cy^2$ \\      
\noalign{\smallskip}
\noalign{\smallskip}
\noalign{\smallskip}
\hline
\noalign{\smallskip}
i & $b_{i}$ & $b_{i}$   & units  \\
& & IERS 2010 & \\
\noalign{\smallskip}
\hline 
\noalign{\smallskip}
0 & $ -0.0036 \pm 0.0006$  &  $ -0.006951 $ & $''$  \\     
1 & $ -0.024 \pm 0.006 $ & $ -0.025896  $ & $''/cy$  \\
2 &  $ -22.539 \pm 0.031$  &  $ -22.4072747 $ & $''/cy^2$ \\      
\noalign{\smallskip}
\hline
\end{tabular}
\tablefoot{Estimates of the parameters for the largest polynomial terms in the Eq.~(\ref{Eq4}) for $X_{IRP}(t)$ and $Y_{IRP}(t)$ and for the IAU2006/ 2000A development; Eq. (5.16) in \cite{iers10}.}
\end{table}

\begin{table}[!h]
\caption{Coefficients of the forced nutation model. }
\label{table2}
\tiny
\begin{tabular}{c|cc|cc} 
\hline
\noalign{\smallskip}
Period, & $A^C_{i}$ & $A^S_{i}$ &  $B^C_{i}$  &  $B^S_{i}$  \\
\noalign{\smallskip}
year& mas & mas & mas & mas   \\
\noalign{\smallskip}
\hline 
\noalign{\smallskip}
0.037402 &  $-70.0 \pm 0.6$  &  $ -43.7 \pm 0.6$    &   $  48.0 \pm 0.5 $  &  $-77.8 \pm 0.5$ \\     
0.3333 &  $5.7 \pm 0.6 $ & $16.7  \pm  0.6 $
& $-20.4 \pm 0.5 $ & $6.6 \pm 0.5 $
\\ 
0.4872 & $ 6.0 \pm 0.6 $  &   $ -0.1 \pm 0.6 $  &    $  -0.8  \pm 0.5 $ &  $ 7.6 \pm 0.5 $      \\
0.5 & $178.6 \pm 0.6 $  &   $487.8 \pm 0.6 $  &    $  -535.0  \pm 0.5 $ &  $ 194.9 \pm 0.5 $      \\ 
1.0 &  $-6.0 \pm 0.6 $  &  $ 46.7 \pm 0.6 $  &  $  15.3 \pm 0.5 $ &   $  -3.8  \pm 0.4 $ \\
4.01 & $ -11.5 \pm 0.6 $  &   $  11.1 \pm 0.7 $  &    $  0.4  \pm 0.5 $ &  $ -0.5 \pm 0.5 $      \\
9.3 & $-77.9 \pm 0.6$  & $ 28.5 \pm 0.6 $     & 
    $  28.4 \pm 0.5    $   &    $ 83.7 \pm 0.5 $  \\
18.6 & $-5608 \pm 1$  &  $ -3927 \pm 1$ &  $-5294 \pm 1 $  &  $ 7533 \pm 1 $ \\
\noalign{\smallskip}
\hline
\end{tabular}
\tablefoot{Amplitudes for eight periodical harmonics in Eq~(\ref{Eq4}). Coefficients $A^C_{i}$ and $A^S_{i}$ are related to $X_{IRP}$, and coefficients $B^C_{i}$ and $B^S_{i}$ are related to $Y_{IRP}$ variations, correspondingly.}

\end{table}

\begin{figure} 
	\centering
    \includegraphics[width=0.45\textwidth]{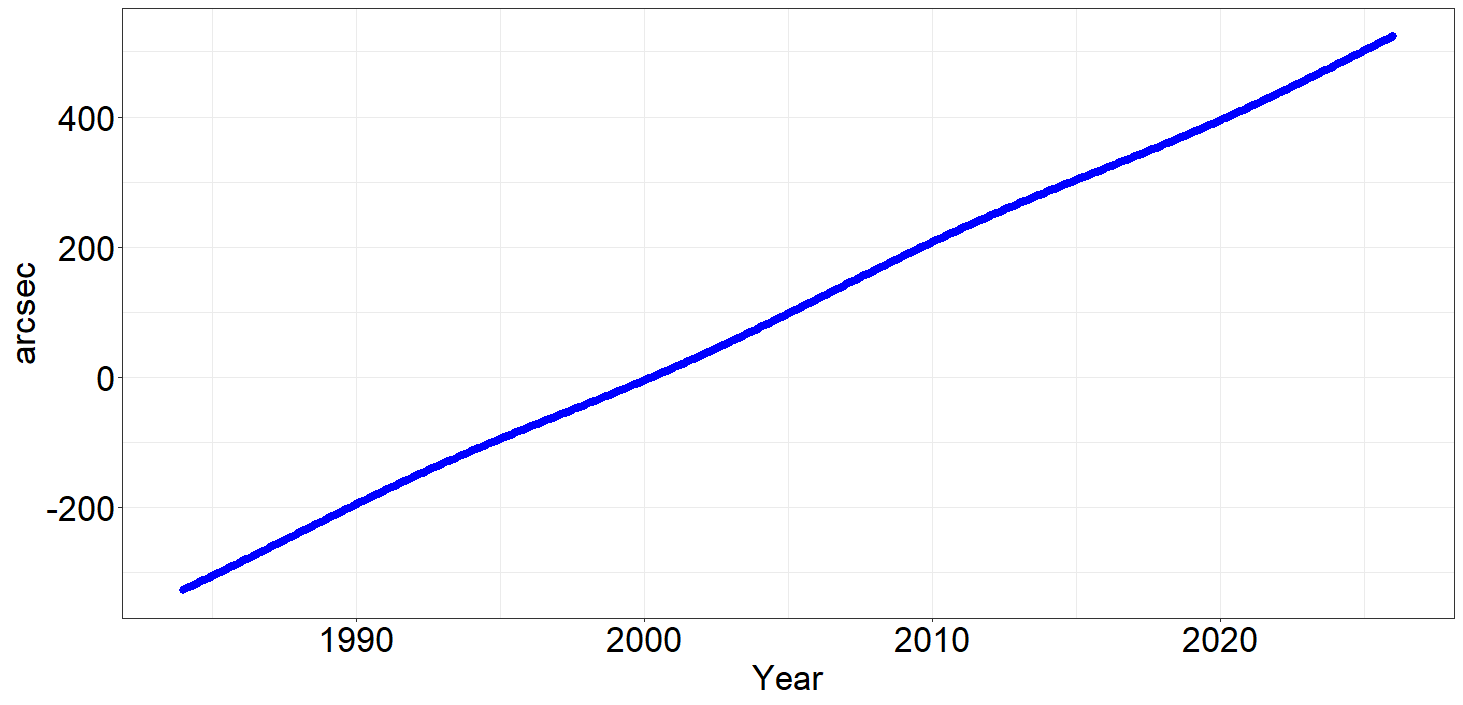}
	\includegraphics[width=0.45\textwidth]{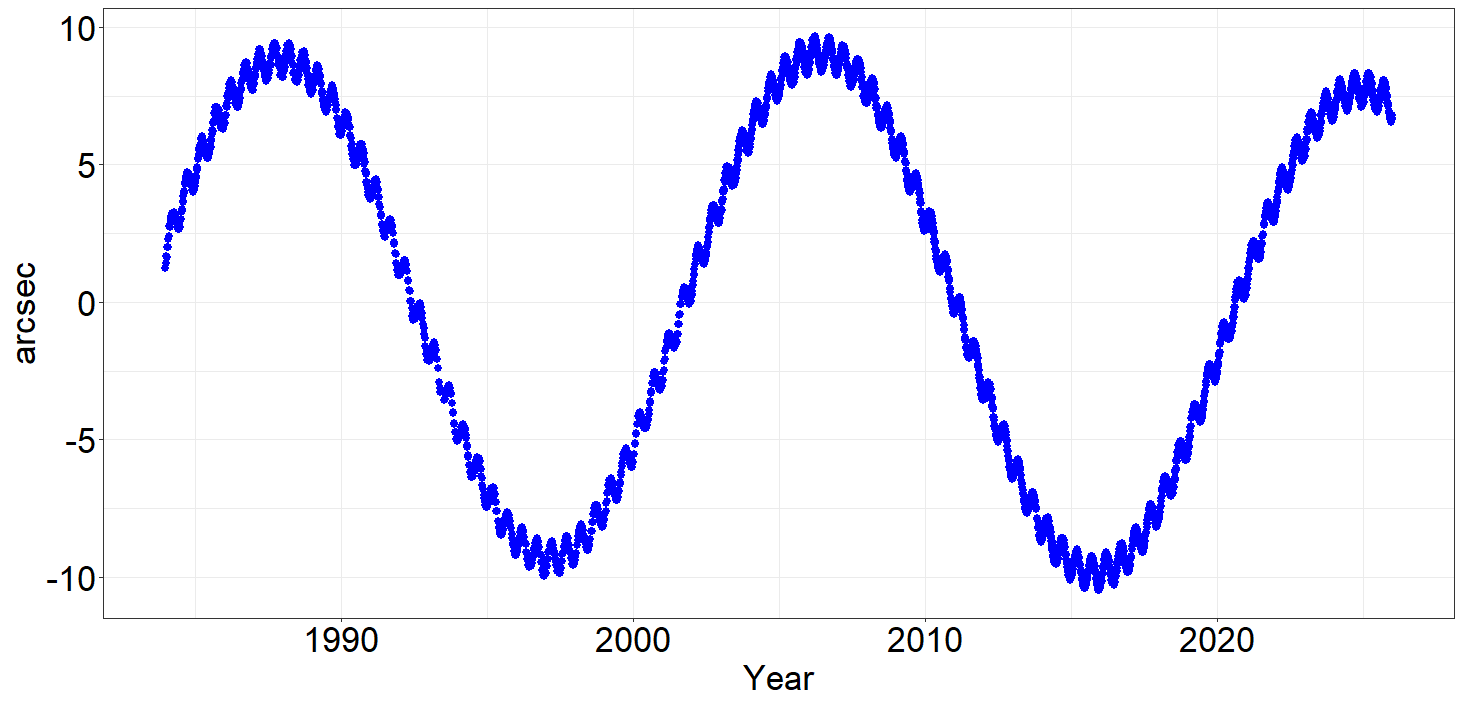}  
	\caption{Time series of the IRP directions $X_{IRP}$ and $Y_{IRP}$ on the upper and lower panels, respectively, in 1983-2026 calculated by Eq.~(\ref{Eq2})}
\label{fig1} 
\end{figure}

\begin{figure} 
	\centering
    \includegraphics[width=0.45\textwidth]{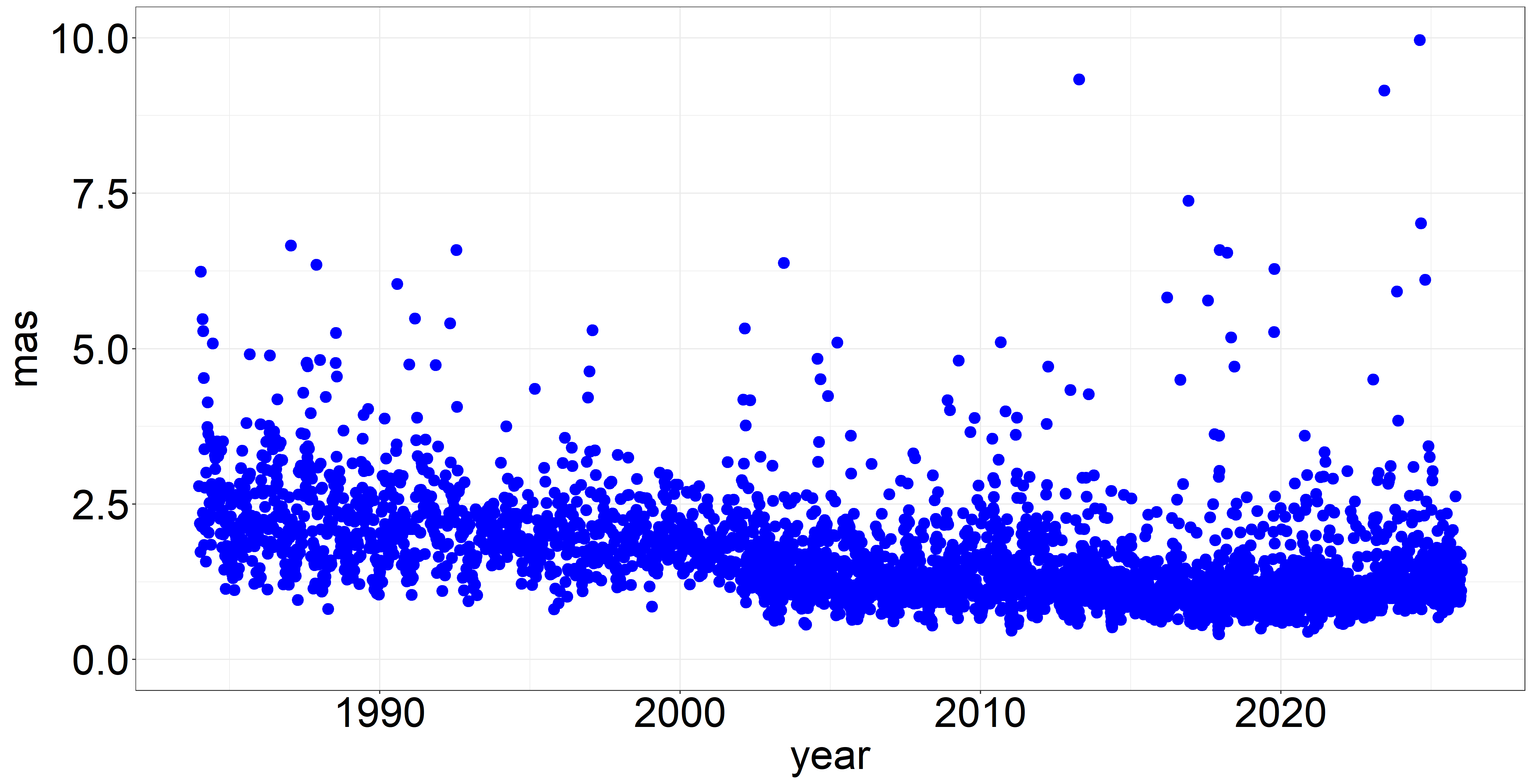}
	\includegraphics[width=0.45\textwidth]{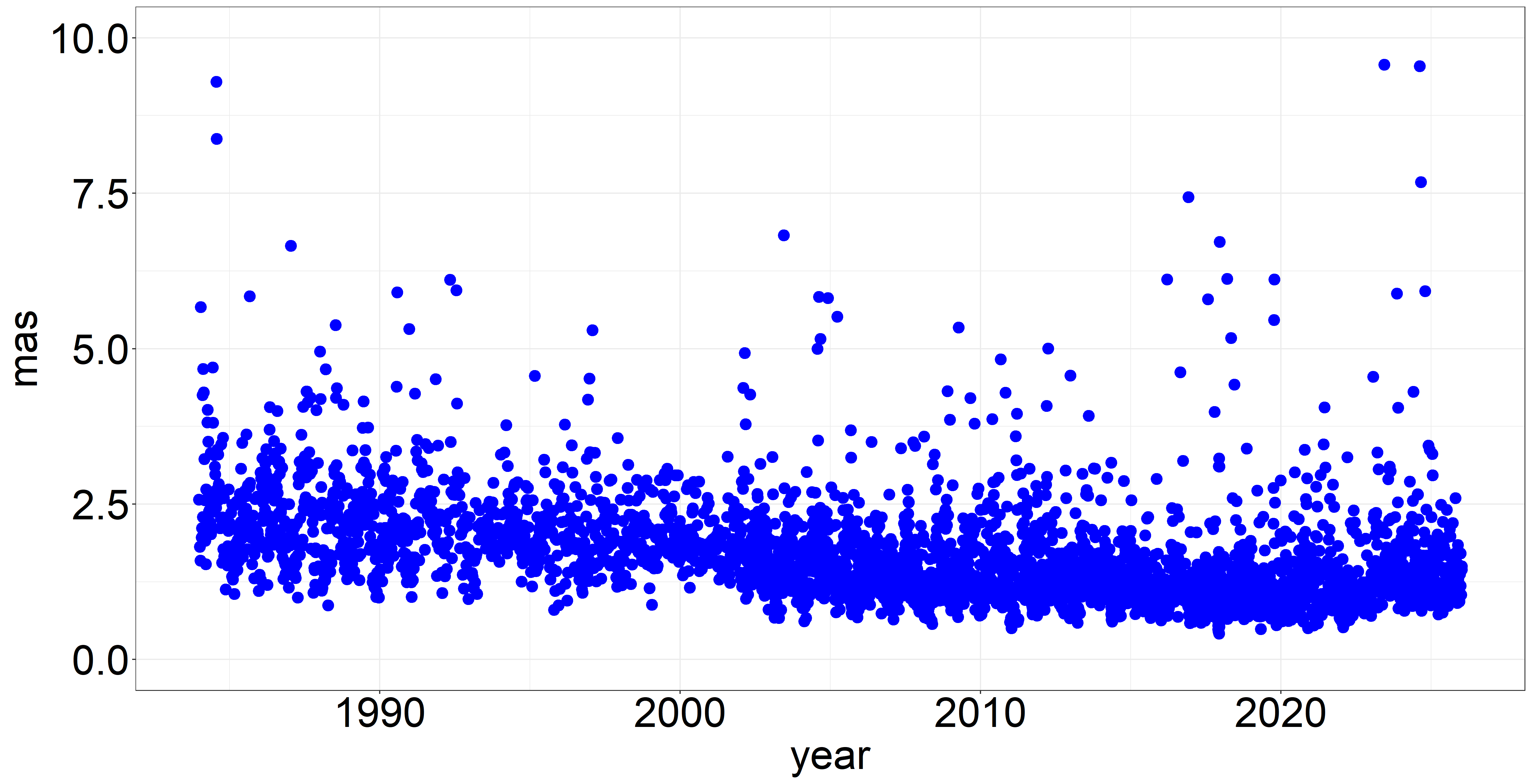} 
	\caption{Formal errors  $\sigma_{X_{IRP}}$ and $\sigma_{Y_{IRP}}$ on the upper and lower panels, respectively, in 1983-2026 calculated by Eq.~(\ref{Eq3})}
\label{fig1a} 
\end{figure}

The time series $X_{IRP}$ and $Y_{IRP}$ were approximated with a combination of the non-period polynomials and harmonic signals for the epoch $T$ (e.g. \cite{iers10}, \cite{Capitaine2008}). For a quick approximation, we used the eight most powerful periodic signals: $P_{1}$= 0.037402 year (13.661  days), $P_{2}$ = 0.3333 year, 
$P_{3}$ = 0.4872 year,
$P_{4}$=0.5 year, $P_{5}$=1.0 year,
$P_{6}$ = 4.01 year,
$P_{7}$=9.3 year, and $P_{8}$=18.6 years.
The following equation presents the model for the $X_{IRP}$ and $Y_{IRP}$ time series:

\begin{equation}
\begin{aligned}
X_{IRP}(t) = \sum^2_{i=0} a_{i}\Delta t^i +  \sum^8_{i=1} (A^c_{i}\cos\frac{2\pi\Delta t}{P_{i}} + A^s_{i} \sin\frac{2\pi\Delta t}{P_{i}})  \\
Y_{IRP}(t) = \sum^2_{i=0} b_{i}\Delta t^i +  \sum^8_{i=1} (B^c_{i}\cos\frac{2\pi\Delta t}{P_{i}} + B^s_{i} \sin\frac{2\pi\Delta t}{P_{i}}),
\end{aligned}
	\label{Eq4}   
\end{equation}
where $X_{IRP}$ and $Y_{IRP}$ are in arcseconds and $\Delta t$ is in centuries for the long-term trends from Table~\ref{polynom} ($\Delta t = \frac{ T -2000.0}{100}$) and in years for the harmonic signals from Table~\ref{table2} ($\Delta t = T -2000.0$). 

The estimate of the angle $\theta$ is $2004.182 \pm 0.077 $ $''$/cy in Table \ref{polynom}, and it may be converted to the precession constant value as $\psi_{A} = \frac{\theta_{A}}{\sin\epsilon_{0}}$. For $\epsilon_{0} = 84381.406$$''$, this is equal to $\psi_{A}= 5038.457  \pm 0.019 $$''$/cy, which refers to the mean observational interval 2004.0, and by considering the linear change of $\psi_{A}$ on approximately $-1.07$~ $''$/cy, one could reckon that this is equivalent to $\psi_{A}= 5038.500 \pm 0.019''$/cy for the reference epoch 2000.0. }This new estimate is consistent with the official IAU2006A value of $\psi_{A}= 5038.4815 $ $''$/cy.

Major standard nutation harmonics have been found in the spectrum of the Earth's rotational velocity, and their estimates are also presented in Table~\ref{table2}.
In addition, we detected a strong peak with a 4-year period on prograde and retrograde domains (see Figure~\ref{fig2}). The amplitudes of the signal in the prograde and retrograde domains are close, so it may appear in a form of elongated ellipse with an amplitude of $16.0 \pm 0.6$ mas.
Although the nature of this signal's origin is still unknown, it was added to the model through Eq~(\ref{Eq4}) because its amplitude significantly exceeds the formal error. Apparently, spectral analysis of the post-fit residuals has not been done in other studies, so it is hard to say whether the presence of this signal has been confirmed by other authors. The signal itself appears fairly stable in the spectrum and cannot be explained by a combination of some other harmonics. At the same time \cite{Mound_2003} predicted a theoretical 4-year signal in the length-of-day spectrum due to the free oscillation mode of the core-mantle system. However, it is not clear whether this signal is also predicted in the nutation time series.

\begin{figure} 
	\centering
    \includegraphics[width=0.5\textwidth]{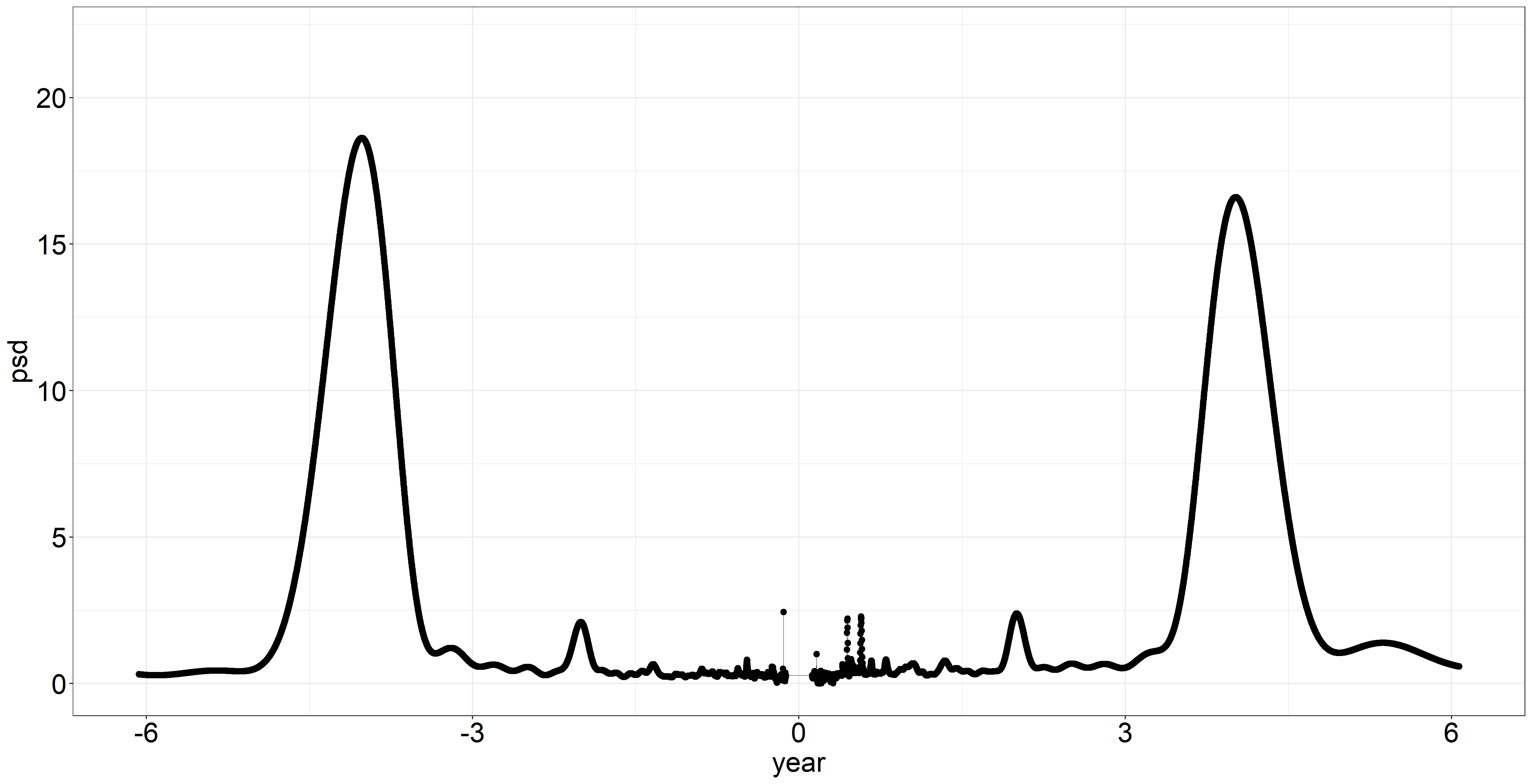} 
	
	\caption{Two-dimensional spectrum of $X, Y$ variations in 1983-2026 calculated after removal of seven terms in  Eq.~(\ref{Eq2}) without a 4-year signal.}
\label{fig2} 
\end{figure}

\subsection{Analysis for the nutation offset from VLBI group delays}

The correction to the precession constant could also be obtained with the geodetic VLBI group delay data. However, the extraction of the linear trend coefficients is not a trivial task due to the effect of the FCN, which dominates after removal of the periodic signals (subsection 3.2; \cite{Capitaine_2009}, \cite{Krasna_2013}, \cite{Gattano_2016}, \cite{Belda2017}).
The FCN effect is hard to model because its period and amplitude are variable. Therefore, we approximated the daily nutation offset time series using model Eqs.~(\ref{Eq5}) and (\ref{Eq6}) with a constant period of the FCN (period $P_{0}$ = 430 days) and amplitude modulation. The choice of this particular signal period value is not tied to any specific reason; it is just convenient for building an approximate model. To represent modulation, we noted that the `disappearance' of the FCN occurs approximately once every 20 years, which may be caused by corresponding changes in climate, the global ocean, or long-term geological factors (\cite{Schlesinger_1994}, \cite{McCarthy_2015}, \cite{Zotov_2016}). Although precise models for length-of-day variations with periods longer than 20 years have not yet been developed due to the insufficient duration of high-precision observations, the authors mentioned above note a significant effect of various processes in the interval from 20 to 60 years.
Therefore, in this work we consider the FCN as a process with an amplitude modulated with a period of 15,000 days (approximately 41 years). As is shown below, such a value of the modulation period leads to a convincing approximation of the observed FCN variations.

Strictly speaking, the time series of the $X, Y$ angles for the FCN are not the same as the $X_{IRP}$ and $Y_{IRP}$ angles because they are obtained from different types of observables. However, as we show in the appendix, the analytical representation of the two transformation matrices is very close, and the expected difference between the two pairs of parameters is likely to be negligible.

 \begin{table*}[!h]
\caption{Parameters of the FCN model.}  
\label{gelya}
\begin{tabular}{cccccccc}
\noalign{\smallskip}
\hline
\noalign{\smallskip}
Coeff & $dX$ & $dY$ & $dY$  &   $dX_{J2}$ & $dY_{J2}$ & $dY_{J2}$  & units  \\
&& (quadratic) & (cubic) && (quadratic) & (cubic) & \\
\hline  
\noalign{\smallskip}
$A_{0}$ &  $4.0\pm   7.4$ & $-81.5  \pm  7.9 $  & $-74.8 \pm 8.5$ &
  $ 3.6\pm  7.3$ & $-77.8  \pm  7.8 $  & $ -71.5 \pm 8.3$ &
$\mu$as\\
$A_{1}$& $-2.7 \pm 1.3$&  $  -6.6 \pm 1.3$ &   $-2.3 \pm 2.3$
& $-2.9   \pm 1.2$&  $  -6.7 \pm 2.3$ &   $ -2.7  \pm 2.3$
& $\mu$as/year\\
$A_{2}$ &$0.44 \pm 0.05$& $  -0.04 \pm 0.05$&  $-0.7 \pm 0.3$
 &$-0.11   \pm 0.05$& $ -0.04 \pm 0.05$ &   $-0.7 \pm 0.3$
&$\mu$as/year$^2$\\
 $A_{3}$&           &    & $ 0.022  \pm0.009$  
 &             &        &  $0.021 \pm 0.009$       & $\mu$as/year$^3$  
 \\
\noalign{\smallskip}
$L_{1}$ &$-35.0  \pm4.0$& $ 48.4\pm  4.3$&   $56.5 \pm5.6$
&$ -31.0  \pm 4.0$& $ 49.7\pm  4.3$&  $57.5 \pm 5.5$ 
& $\mu$as\\
$L_{2}$ &$55.1 \pm 4.1$&   $-43.23 \pm 4.3$&    $-54.0\pm 6.4$
&$53.8 \pm 4.0 $&   $-42.8 \pm 4.3$& $-53.0 \pm 6.3$
& $\mu$as\\
\noalign{\smallskip}

$D_{1}$ &$43.9  \pm 5.7$ &   $ 18.1  \pm 6.1 $ &   $16.6 \pm6.1$&
$42.0  \pm 5.6$ &   $ 17.4  \pm 6.0 $ & $16.1\pm  6.0 $
&  $\mu$as\\

$D_{2}$ &$7.9  \pm 5.7$ &   $ -30.6 \pm 6.1 $ &   $-30.6 \pm6.1$&
$7.5  \pm 5.6$ &   $ -32.1  \pm 6.1 $ & $-32.1\pm  6.0 $
&  $\mu$as\\

$B_{1}$ &$-9.0  \pm 5.9$ &   $ -82.6  \pm 6.4 $ &   $-83.4 \pm6.34$&
$-35.1  \pm 4.9$ &   $ -84.0  \pm 6.3 $ & $-84.7\pm  6.3 $
&  $\mu$as\\
$B_{2}$ &$101.6 \pm 7.7$ & $  144.4 \pm 8.2$ &$146.26\pm 8.2$
 &$103.1  \pm 7.6 $ & $  145.9\pm 8.1$ &  $147.5\pm 8.1$ &$\mu$as\\ 

$B_{3}$ &$-102.5 \pm 5.6$&$ 35.3  \pm 6.1 $ & $34.8 \pm6.1$
 &$-103.2   \pm 5.6$&$ 35.4  \pm 6.0 $ &$ 34.9 \pm 6.0 $& $\mu$as
 \\
 
$B_{4}$ &$130.1  \pm 7.8 $ & $ -144.5 \pm 8.3$ &    $-144.6 \pm8.3$
&$ 133.1 \pm 7.6 $ & $ -142.6    \pm 8.2$ & $-142.8 \pm 8.2$
&$\mu$as\\
\noalign{\smallskip}
\hline
\end{tabular}
\tablefoot{Estimates of the approximation coefficients from Eq.~(\ref{full}) for $dX$ and $dY$   corrections to the IAU 2006 precession model and to the IAU $2006_{J_{2}}$ precession model (\cite{Liu_2025}). The $dY_{IRP}$ estimates are shown for quadratic and cubic approximation for each precession model.}
\end{table*}

The suggested model for the FCN starts from the basic equation:

\begin{equation}
f(t) = C(t)\Bigl(\cos(\frac{2\pi\Delta t}{P_{0}} + \varphi_{0})\Bigr),\\
	\label{Eq5}  
\end{equation}

where $P_{0} = 430$ days and $C(t)$ is a slowly changing amplitude with a period of $P_{C} = 15000$ days. 

\begin{equation}
C(t) = C_{0} \Bigl(1+ C_{1}\cos(\frac{2\pi\Delta t}{P_{C}} + \varphi_{C})\Bigr),\\
	\label{Eq6} 
\end{equation}

and $\Delta t$ is in years ($\Delta t = T -2000.0$).

A combination of Eqs~(\ref{Eq5}) and (\ref{Eq6}) leads to

\begin{equation}
\begin{aligned}
f(t) & =  C_{0} \Bigl(1+ C_{1}\cos(\frac{2\pi\Delta t}{P_{C}} + \varphi_{C})\Bigr) \cos\!\left(\frac{2\pi\Delta t}{P_{0}} + \varphi_{0}\right) =\\[4pt]
&=C_{0}\cos\!\left(\frac{2\pi\Delta t}{P_{0}} + \varphi_{0}\right) +\\[4pt]
&+C_{0} \Bigl(C_{1}\cos(\frac{2\pi\Delta t}{P_{C}} + \varphi_{C})\Bigr)
\cos\!\left(\frac{2\pi\Delta t}{P_{0}} + \varphi_{0}\right) = \\[4pt] 
&=D_{1} \cos\frac{2\pi\Delta t}{P_{0}} + D_{2}\sin\frac{2\pi\Delta t}{P_{0}}  +\\[4pt]
&+ B_{1}\cos\frac{2\pi\Delta t}{P_{C}}\cos\frac{2\pi\Delta t}{P_{0}} 
   + B_{2}\cos\frac{2\pi\Delta t}{P_{C}}\sin\frac{2\pi\Delta t}{P_{0}} \\
&\quad + B_{3}\sin\frac{2\pi\Delta t}{P_{C}}\cos\frac{2\pi\Delta t}{P_{0}}
   + B_{4}\sin\frac{2\pi\Delta t}{P_{C}}\sin\frac{2\pi\Delta t}{P_{0}} \,,
\label{Eq7}
\end{aligned}
\end{equation}

where $D_{1} = C_{0} \cos\varphi_{0}$, $D_{2} = -C_{0} \sin\varphi_{0}$, and $B_{1} = C_{0}C_{1}\cos\varphi_{C}\cos\varphi_{0}$, 
$B_{2} = -C_{0}C_{1}\cos\varphi_{C}\sin\varphi_{0}$,
$B_{3} = -C_{0}C_{1}\sin\varphi_{C}\cos\varphi_{0}$,
$B_{4} = C_{0}C_{1}\sin\varphi_{C}\sin\varphi_{0}$.

Additionally, the model for the FCN includes the constant offset, linear and quadratic trend, and signal with period $P_{2} = 18.6$ year.

\begin{equation}
\begin{aligned}
f(t) = A_{0} + A_{1}\Delta t + A_{2} \Delta t^2
+ L_{1} \cos\frac{2\pi\Delta t}{P_{2}} + L_{2}\sin\frac{2\pi\Delta t}{P_{2}} 
	\label{Eq8}  
    \end{aligned},
\end{equation}

so the final model to be fitted with the least-squares method is given by

\begin{equation}
\begin{aligned}
f(t) = A_{0} + A_{1}\Delta t + A_{2} \Delta t^2
+ L_{1} \cos\frac{2\pi\Delta t}{P_{2}} + L_{2}\sin\frac{2\pi\Delta t}{P_{2}} + \\
+ D_{1} \cos\frac{2\pi\Delta t}{P_{0}} + D_{2}\sin\frac{2\pi\Delta t}{P_{0}} + \\+
B_{1}\cos\frac{2\pi\Delta t}{P_{C}}\cos\frac{2\pi\Delta t}{P_{0}} +  
 B_{2}\cos\frac{2\pi\Delta t}{P_{C}}\sin\frac{2\pi\Delta t}{P_{0}} + \\
 + B_{3}\sin\frac{2\pi\Delta t}{P_{C}}\cos\frac{2\pi\Delta t}{P_{0}} +  
 B_{4}\sin\frac{2\pi\Delta t}{P_{C}}\sin\frac{2\pi\Delta t}{P_{0}}. 
 \\
	\label{full}  
    \end{aligned}
\end{equation}

The coefficients of decomposition are presented in Table~\ref{gelya}. For the $Y$ component, a cubic trend was also applied; however, it does not differ significantly from the model with a quadratic trend.

\begin{figure} 
	\centering
    \includegraphics[width=0.45\textwidth]{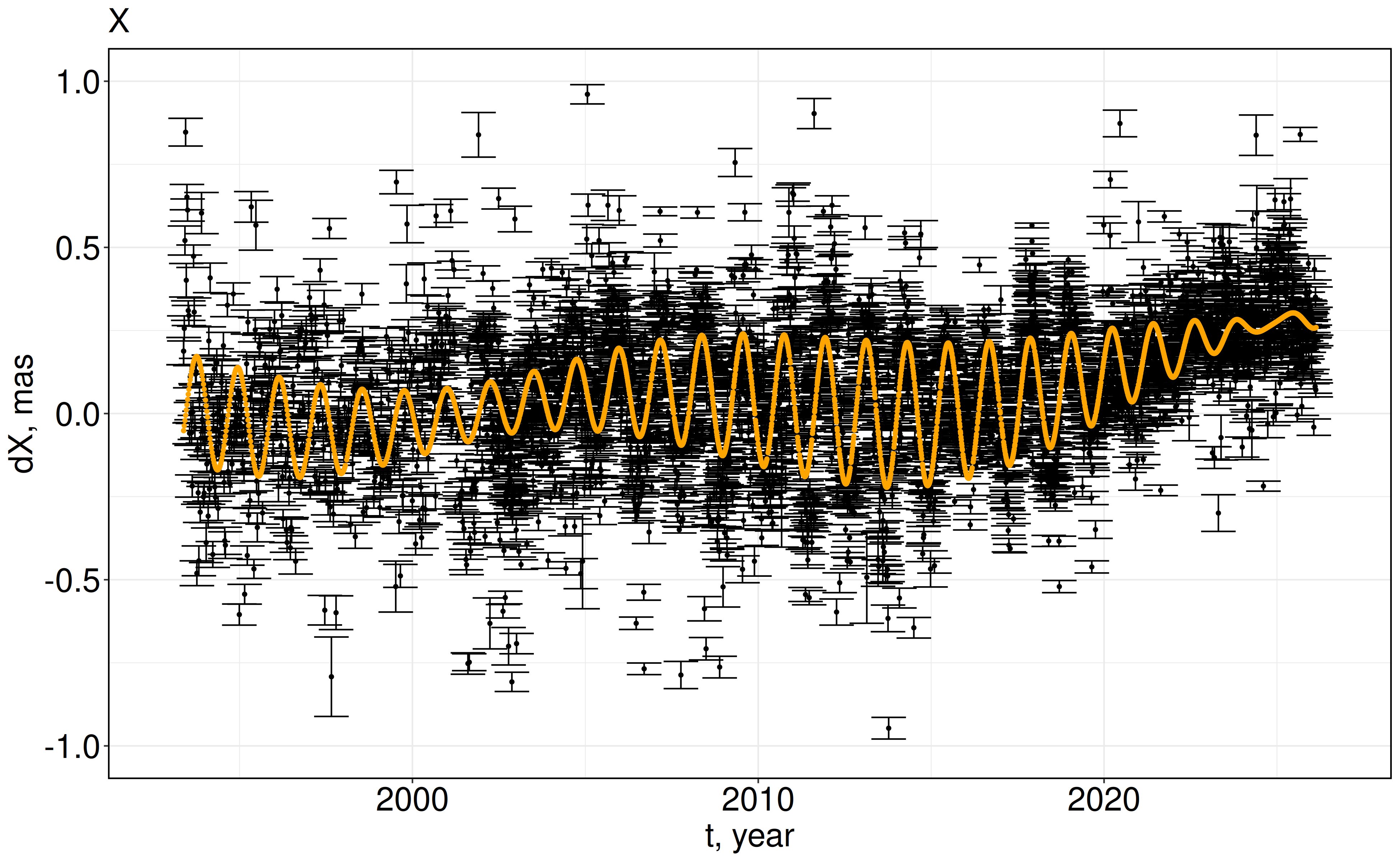}
	\includegraphics[width=0.45\textwidth]{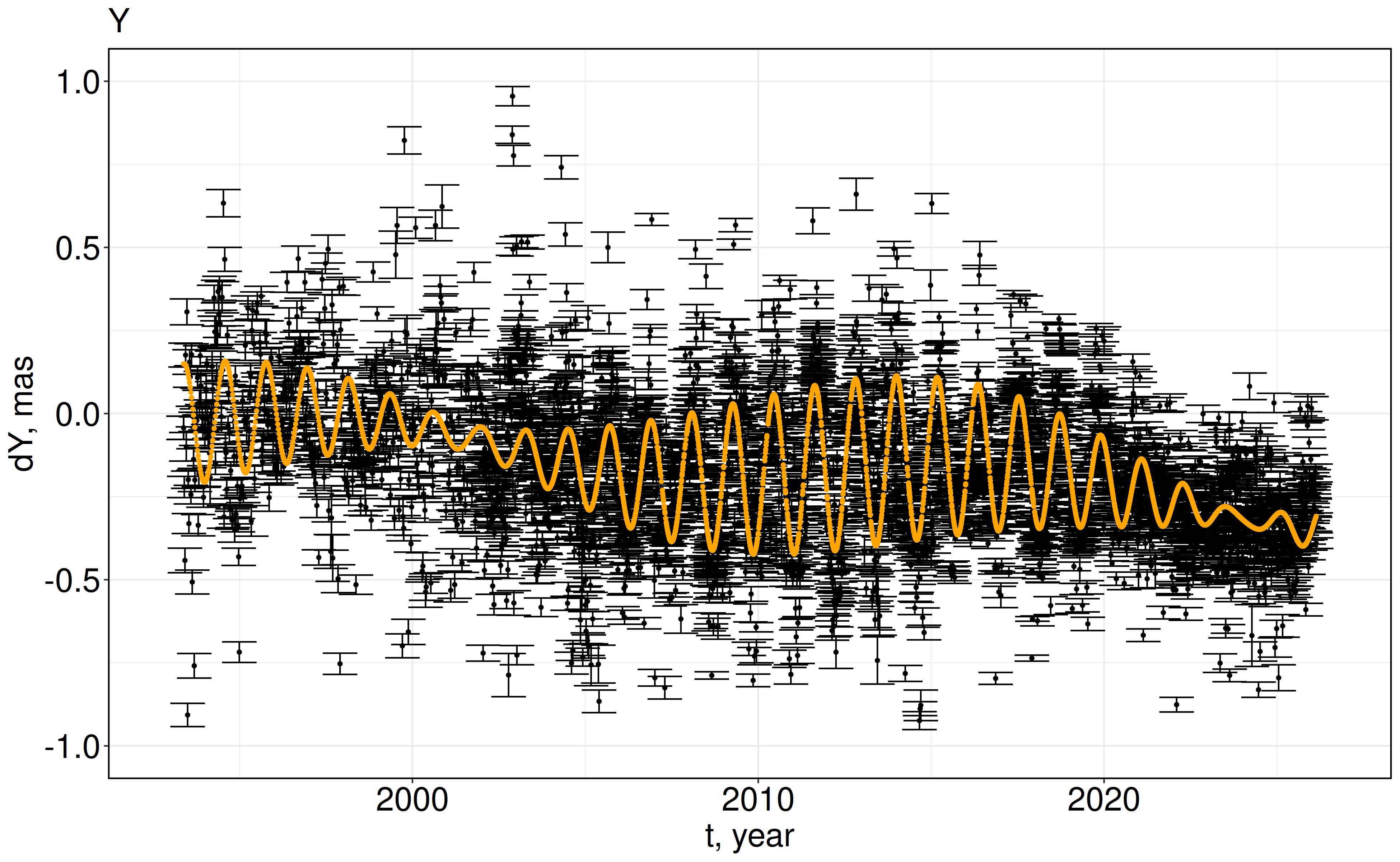} 
	
	\caption{Variations of $dX$ and $dY$ and their approximation with model  Eq~(\ref{full}) that comprise a long-term trend and periodic signal. For $dY$, the model with a quadratic trend is presented here.}
\label{fig3}  
\end{figure}

\begin{figure} 
	\centering
    \includegraphics[width=0.45\textwidth]{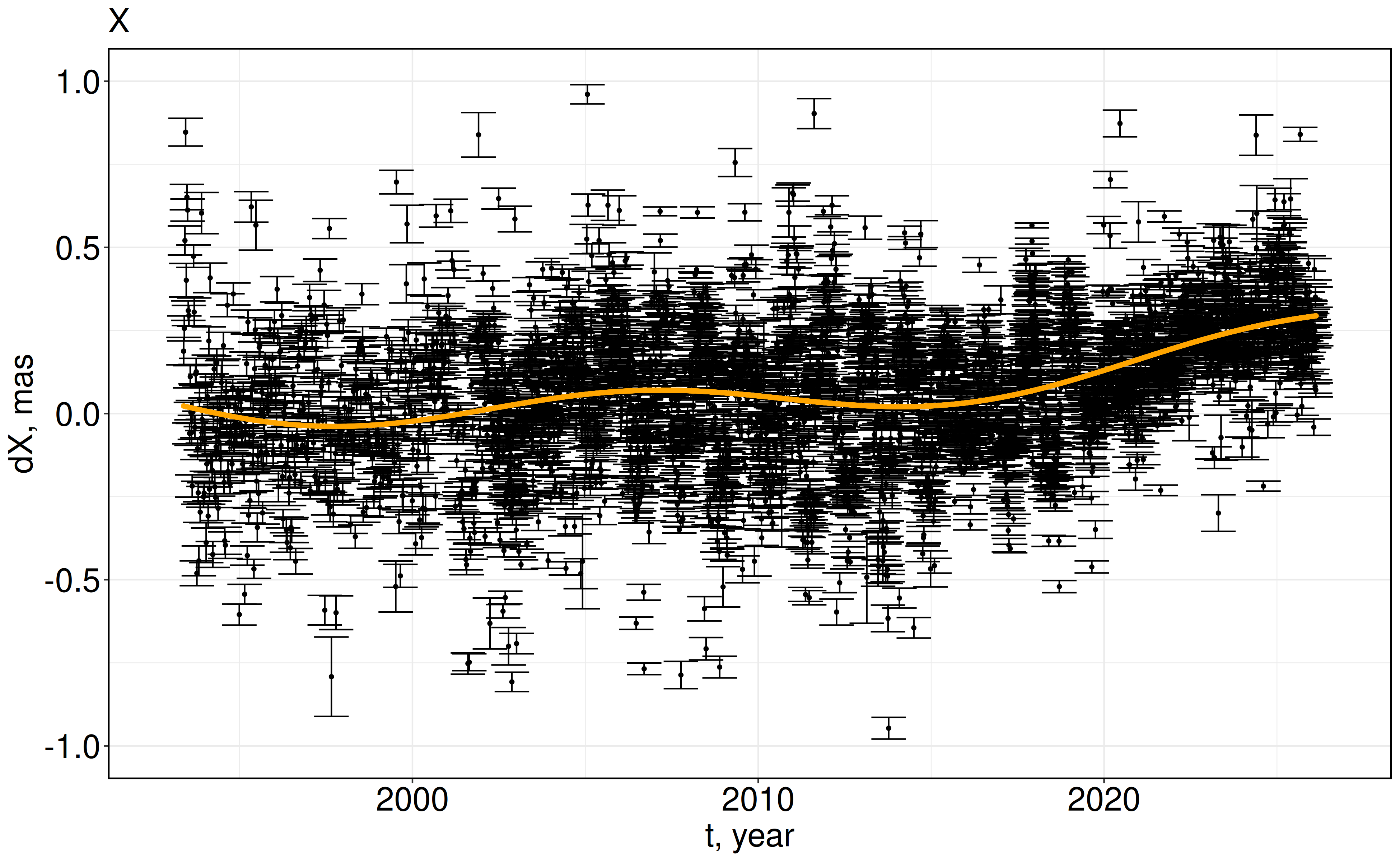}
	\includegraphics[width=0.45\textwidth]{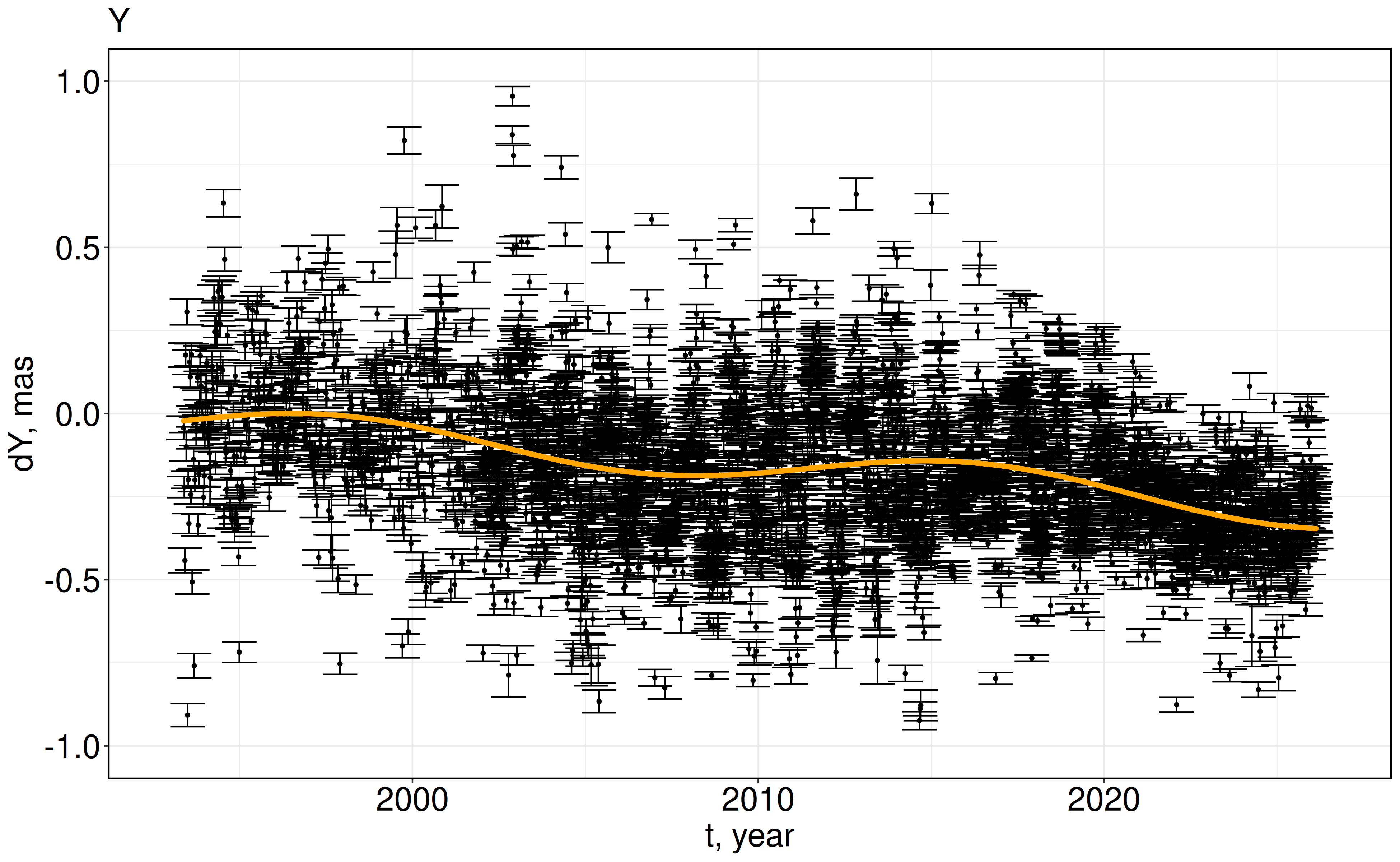}  
 \caption{ Variations of $dX$ and $dY$ and their approximation with  Eq~(\ref{Eq8}) that show only the long-term trend of both component estimates. For $dY$, the model with a quadratic trend is presented here.}
\label{fig4}  
\end{figure}

It should be noted here and until the end of this subsection that in its current form (Eq~(\ref{full})) is designed for approximation of a deterministic process, while the FCN is known to be a stochastic process. That is why the primary purpose of Eq~(\ref{full}) is an approximation of the FCN for, say, operational service of the International VLBI Service (IVS) intensive experiments, so physical interpretation of the parameters in Table~\ref{gelya} should be done carefully.

\subsubsection{Conventional IAU 2006 precession model}

While Eq~(\ref{Eq8}) shows only the long-term trend of the celestial pole offset, the approximation proposed in Eq~(\ref{full}) combines this long-term trend with the periodic variations of the FCN. 
As shown in Figures \ref{fig3} and \ref{fig4}, such a model effectively approximates the variations of the FCN with the `disappearance' of the signal around 2002 and 2023. Although this model is rather rough, this simplification allows one to get rid of the cumbersome tabulation of the FCN as it is adopted by the IERS standards (\cite{iers10}). The model, though, also incorporates linear and quadratic effects, which are shown in Figure~\ref{fig4} without the periodical systematic. The fitting of the time series with the quadratic model produces the coefficients of the long-term trend that could be attributed as correction to the theoretical precession model (Table~\ref{gelya}). 
The linear trend correction to the precession model is not significant statistically, but  
the quadratic trend is quite significant, as can be seen in Table~\ref{gelya}, and it produces an apparent deviation from zero in Figure~\ref{fig4}. 

The approximation proposed in Eq~(\ref{full}) reflects the long-term and periodic variations of the FCN, while Eq~(\ref{Eq8}) shows only the long-term trend. This trend is caused by the Earth's global dynamic evolution. In the next subsection, we demonstrate the impact of the Earth's flattening instability on the trend. Such an approximation (Eq~\ref{Eq8}) is useful for visualisation of the long-term effect, which results in a significant deviation of the two components from zero by 2025.

\subsubsection{Conventional IAU 2006 precession model with a $J_{2}$ correction}

\cite{Liu_2025} developed a correction to the IAU 2006 model coefficients due to a quadratic trend in the $J_{2}$ of Earth's dynamic flattening parameter. In accordance with the new modeling of the parameter $J_{2}$ , the deviation of the $X$ angle from zero is essentially reduced (upper panels of Figure~\ref{fig5} and Figure~\ref{fig6}); however, the correction to the $Y$ component is not noticeable because variations of $J_{2}$ have almost no effect on $Y$. 

\cite{Liu_2025} concluded that the impact of the additional correction is negligible for the component $Y$ but significant for the component $X$. Our comparison of estimates in Tables \ref{gelya} confirms this conclusion. The estimates of the nutation coefficients diverge by up to 4 microseconds for the $X$ component, and the estimates of the linear and quadratic trends also have a noticeable discrepancy, resulting in the $X$ graphs in Figure~\ref{fig4} and Figure~\ref{fig6}  being significantly different.

\begin{figure} 
	\centering
    \includegraphics[width=0.45\textwidth]{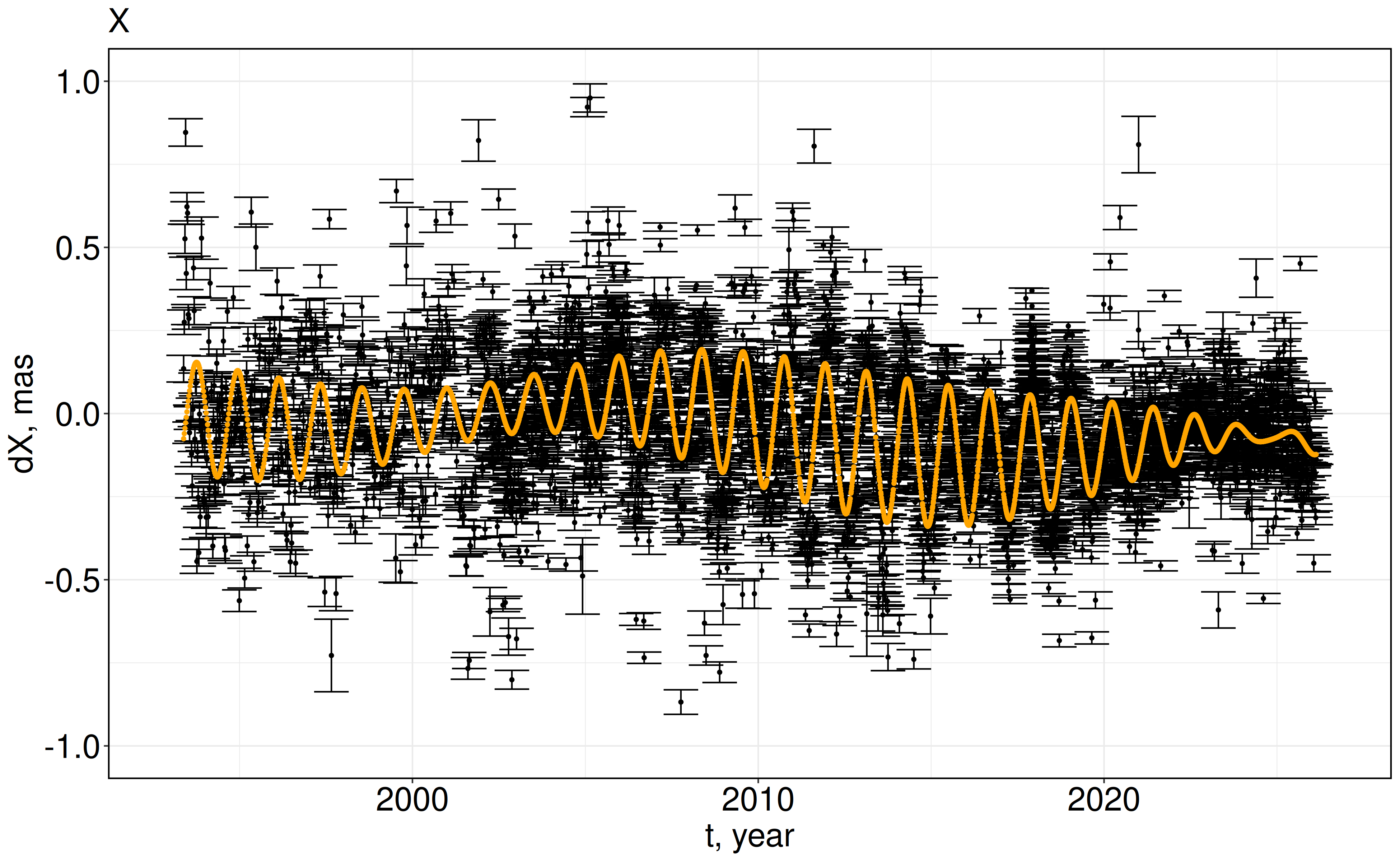}
	\includegraphics[width=0.45\textwidth]{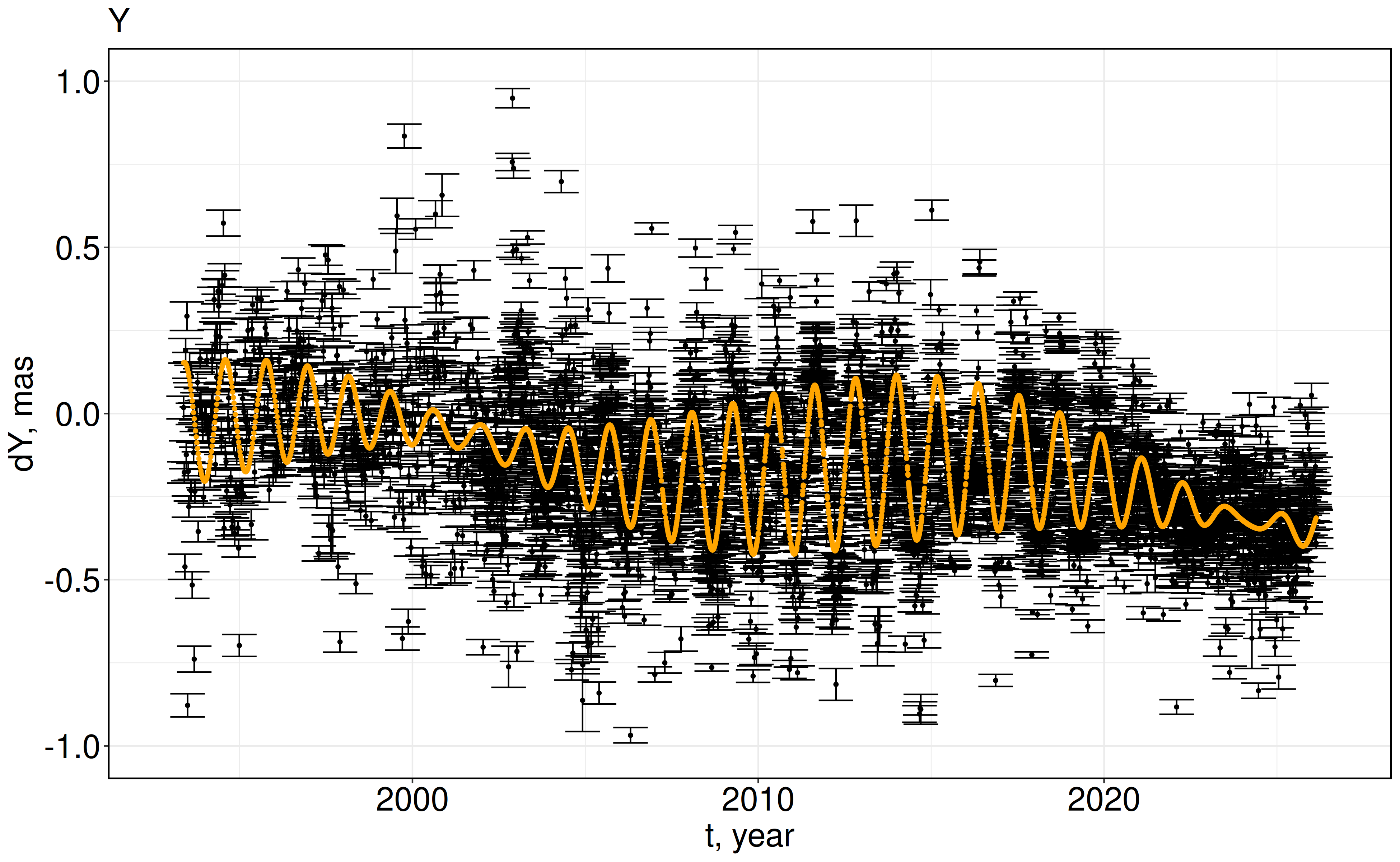}  
	\caption{Variations of $dX$ and $dY$ after correction for $J_{2}$ and their approximation with model  Eq~(\ref{full}). For $dY$, the model with a quadratic trend is presented here.}
\label{fig5} 
\end{figure}

\begin{figure}  
	\centering
    \includegraphics[width=0.45\textwidth]{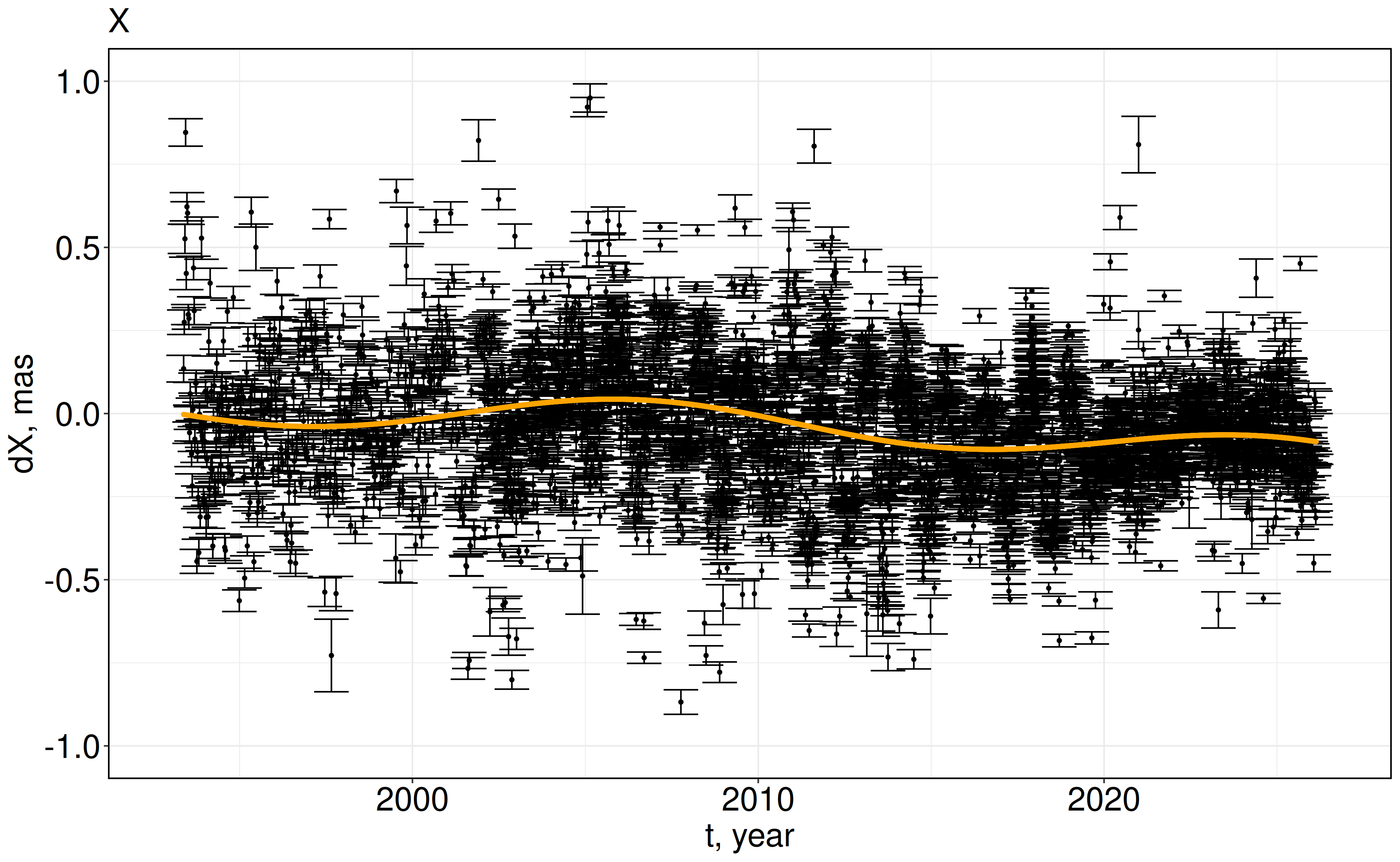}
	\includegraphics[width=0.45\textwidth]{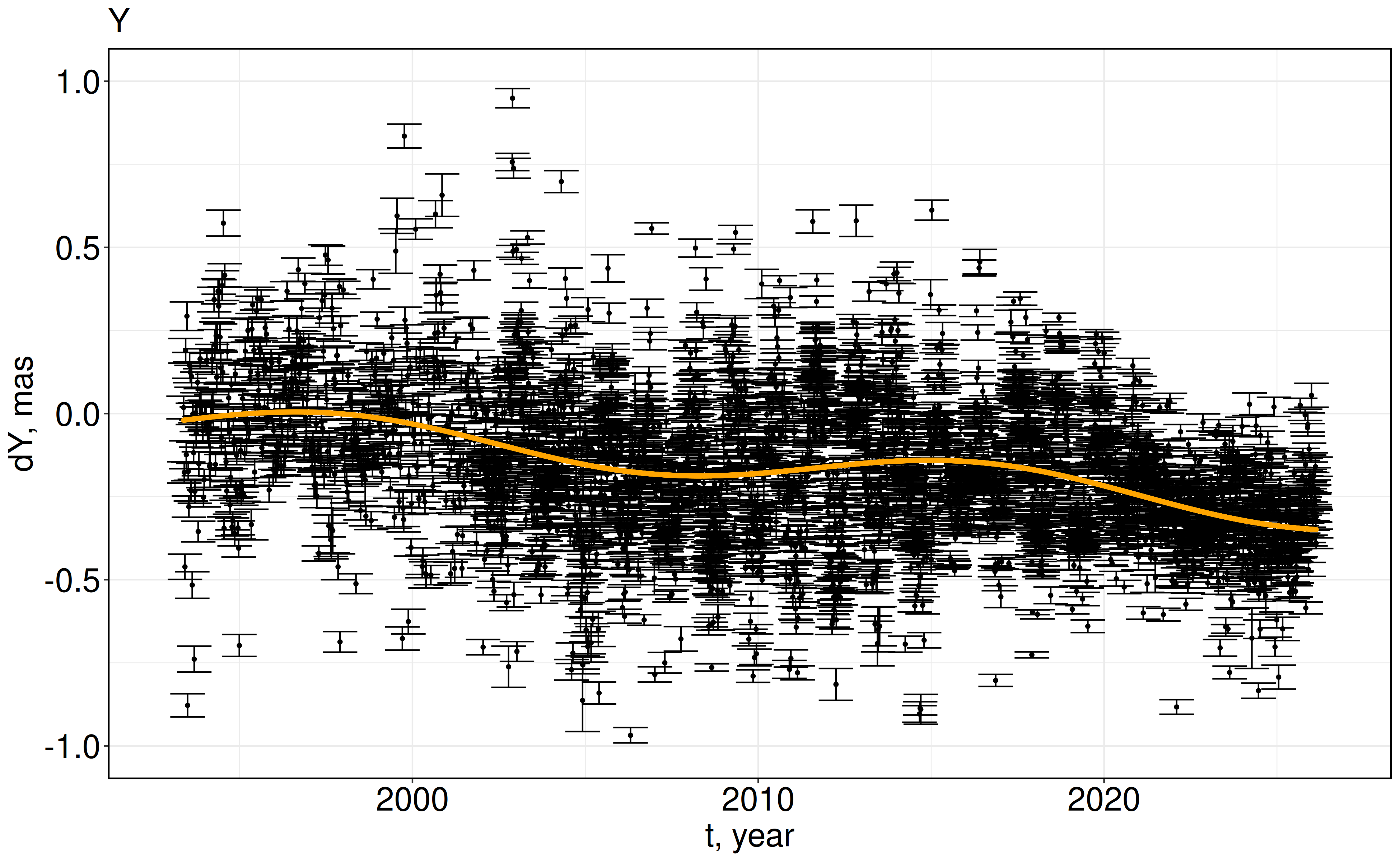}  
\caption{Variations of  $dX$ and $dY$ after the correction for $J_{2}$ and their approximation with model Eq~(\ref{Eq8}). For $dY$, the model with a quadratic trend is presented here.}
\label{fig6}  
\end{figure}

\subsection{Transformation from ITRS to GCRS}

The IERS 2010 Conventions model (\cite{iers10}) demands an introduction of two angles $(E, d)$ that define the CIP position in the GCRS, and the angles $(X, Y)$ are calculated as $X = \sin d\cos E$, $Y = \sin d\sin E$. A new definition of the angles $X_{IRP}$ and $Y_{IRP}$ based on the observational results in accordance with Eq~(\ref{Eq2}) necessitates a new transformation between ITRS and GCRS.

The development of the transformation matrix is shown in the Appendix. As intuitively expected from a visual comparison of matrix $Q(t)$ from the IERS 2010 Conventions and $Q_{IRP}(t)$ (Eq~\ref{B_2}), the deviation from the standard transition matrix (e.g. \cite{Wallace_2006}, \cite{Capitaine_2009} ) turned out to be minimal and boiled down to a discrepancy in a small parameter related to the product of the $X$ and $Y$ angles.
The official IAU2006 model includes the so-called CIO locator to define the position of the CIO that needs a numerical integration -- Eqs (5.8) and (5.9) in the IERS Conventions 2010 (\cite{iers10}). The corresponding elements of the final transformation matrix, Eq (5.10) in the IERS Conventions 2010, form a symmetric pair of elements, thus violating the common notion that the transformation matrix is a purely rigid rotation matrix.

The newly derived `IRP locator' is given by (\ref{arguments1}), and the two corresponding elements of the $Q_{IRP}(t)$ are not equivalent to the conventional definition. While the lower diagonal element of the $Q_{IRP}(t)$  (\ref{B_3}) is equal to $S$, the upper diagonal element is equal to $S - \frac{X_{IRP}Y_{IRP}}{2}$, thus combining of two transformations: rotation and deformation. The matrix $Q_{IRP}(t)$ in (\ref{B_3}) apparently splits into the rotational (anti-symmetrical) $Q'_{IRP}(t)$ and deformation (symmetrical) $Q''_{IRP}(t)$ matrices.

The rotation matrix $Q'_{IRP}(t)$ contains a similar set of off-diagonal elements is linked to the matrix of angular velocity $\Omega$ made of the three components of the rotation vector (Eq~(\ref{Eq1})) because they origin from the equation

\begin{equation}
\Omega \approx  \dot Q'_{IRP}(t)
\end{equation}

and, respectively, we could introduce a new matrix $D$ to describe the first derivatives of the deformation matrix $Q''_{IRP}(t)$ as follows

\begin{equation}
D \approx \dot Q''_{IRP}(t)
\end{equation}

So in the matrix form,

\begin{equation}
\Omega =
\begin{pmatrix}
  0& -\omega_{z}  &  \omega_{y} \\
  \omega_{z}  & 0 & -\omega_{x}\\
  -\omega_{y} & \omega_{x} & 0
	\label{matrix_rotation}  
\end{pmatrix}  
\end{equation}

\begin{equation}
D =
\begin{pmatrix}
  0& d_{z}  &  d_{y} \\
  d_{z}  & 0 & d_{x}\\
  d_{y} & d_{x} & 0
	\label{matrix_deformation}  
\end{pmatrix}  .
\end{equation}

This leads to the appearance of the symmetric deformation matrix as an addition to the anti-symmetric rotation matrix to describe the geocentric velocity of the VLBI stations (\cite{Titov_2025}).

\begin{equation}
\begin{aligned}
   \frac{\Delta f_{21}}{f}=-\frac{(w_{2}\cdot s)-(w_{1}\cdot s)}{c}
    = -\frac{([(\Omega+D)\times b_{21}]\cdot s)}{c} = \\
    -\frac{([s\times b_{21}]\cdot (\Omega+D)}{c}, 
	\label{velocity}  
\end{aligned}
\end{equation}

where $w_{1}$  and  $w_{2}$ are vectors of the geocentric velocities of the radio telescopes, with geocentric positions assigned by vectors $r_{1}$  and  $r_{2}$; $s$ is the unit vector in the direction of the radio source; and $c$ is the speed of light.

While matrix $\Omega$ describes the well-known rotational velocities affecting the observed frequency change, matrix $D$ represents small velocities associated with the tiny deformation of an elementary spherical trapezoid formed by parallels and meridians on the Earth's surface. This is a mathematical artefact that arises from the non-commutability of finite turns (second-order terms of smallness, such as the '$XY$'). In the transition from finite turns to infinitesimal ones, these second-order terms formally form a symmetric matrix in Eq~(\ref{matrix_deformation}). The symmetric matrix of type $Q''_{IRP}(t)$ is implicitly included in the general IAU2006 transformation from ITRS to GCRS; however, its elements have so far remained outside the scope of research. Theoretically, all three non-diagonal elements of matrix $D$ could be estimated from VLBI observations with Eq~(\ref{velocity}), although it is not clear whether the measured velocities will reflect any real physical processes or if they will turn out to be some kind of fictitious values (e.g. coordinate terms). A detailed analysis fo the matrix $D$ elements will be developed in a follow-up publication.

\section{Conclusion}

In this article, we have shown that despite being ignored for a long time, the VLBI delay rates are very useful for estimating the three components of Earth's angular rotation over a short period of observations. The benefit of calculating the Earth's absolute rotation speed was discussed earlier \citep{Titov_2025}, and in this work we have shown how one can get the direction of the Earth's IRP in the GCRS system from these three components. 
The development of the analytical transition in Eqs~(\ref{Eq2}) and (\ref{Eq3}) led to the time series in Figure \ref{fig1}, which presents a direct projection of the Earth's rotation axis on the celestial sphere in a form of two angles, $X_{IRP}$ and $Y_{IRP}$. The approximation of the track by a simple model provides the empirical precession parameters in Table~\ref{polynom}. Additionally, the new series were used to evaluate the main harmonics of the Earth's axis nutation independently.

The position of the IRP based on the VLBI delay rate may be used for transformation between the ITRS and GCRS, as it works as a `natural' IRP and does not need an additional theoretical construction. If the new procedure demonstrates its effectiveness against the conventional approach, in our opinion, it may lead to a simplification of the process of the transformation between the ITRS and GCRS. Therefore, the advantage of using VLBI delay rate observations lies not in the fact that they are more accurate than the VLBI group delays but in that they allow for a simpler conceptual description of the transformation from ITRS to GCRS.

The precession rate estimate is equal to $\psi_{A}= 5038.500 \pm 0.019 $ $''$/cy for the reference epoch 2000.0 when using the VLBI delay rate data from 1993 to 2026. This rate is consistent with the official IAU2006 value of $\psi_{A}= 5038.4815 $ $''$/cy. Considering the formal error, it can be said that this value agrees with the officially accepted constant of precession within the $1\sigma$ range. Incorporation of the recently studied effect of the $J_{2}$ variations does not change the precession constant, but it may affect the quadratic trend in the more detailed mathematical approximation.

We evaluated the amplitudes of the main nutation harmonics (Table \ref{table2}). In addition, we also discovered a previously unknown signal with a 4-year period and an amplitude of $16.0 \pm 0.6$ mas  in the spectrum of the $X_{IRP}$ and $Y_{IRP}$  time series (Figure \ref{fig2}).

Importantly, the VLBI delay rate is not sensitive to the FCN due to its lower accuracy. Therefore, VLBI group delay data are needed to complete the nutation matrix modelling process.
In one solution, a combination of the VLBI group delay and delay rates looks promising.   

The currently adopted procedure for tabulating the FCN amplitude with a certain time step is inconvenient since it updates with some delay and does not allow for effectively extrapolating this effect several months ahead. Therefore, to fully complete the reduction process, a simple model was developed with a constant period of 430 days and a variable amplitude in the form of a harmonic signal with a period of 15,000 days. While this model is far from perfect, it allowed for a satisfactory approximation of the FCN variation series (Figures \ref{fig5} and \ref{fig6}).

Finally, based on a new model for calculating precession and nutation parameters, we proposed a new method for transformation between GCRS and ITRS, that relies on observational data only and does not require introducing redundant mathematical constructions (Appendix A.2). The new locator $S$ that presents a motion of the IRP with respect to the GCRS origin is does not require an integration for calculating in Eq~(\ref{S1}).

\begin{acknowledgements}
 The authors acknowledge use of VLBI data obtained within the framework of the International VLBI Service for Geodesy and Astrometry (IVS). We are grateful to Dr. Christian Bizouard from Paris Observatory and Dr. Jia-Cheng Liu from Nanjing University for comments and interesting discussion, and express special thanks to the anonymous reviewer of this manuscript. This work is funded by Chinese Academy of Sciences President's International Fellowship Initiative (PIFI), Grant No. 2027PVA0069.

\end{acknowledgements}

\bibliographystyle{aa}
\bibliography{titov}

\begin{appendix}  

\section{Precession and nutation}
\subsection{Traditional approach}

The classical concept the transformation linking the ITRS and GCRS is given by a matrix $A(t)$

\begin{equation}
[GCRS] = A(t)[ITRS],
\label{fig:Eq6}
\end{equation}

where $A(t)$ is a combination of the five matrices:

\begin{equation}
  A(t) =  P(t)N(t)R_{3}(-GST)R_{2}(x_{p})R_{1}(y_{p}).\\
	\label{trans0} 
\end{equation}

Here, $R_{1}, R_{2}, R_{3}$ are three traditional rotation matrices; $x_{p}, y_{p}$ are the components of pole motion; and `GST' is the Greenwich Siderial Time. 
In the traditional concept, the combination of precession and nutation is a decomposition of $PN$ into several rotation matrices:

\begin{equation}
  PN = R_{3}(\zeta_{A})R_{2}(-\theta_{A})R_{3}(z_{A})R_{1}(-\epsilon_{A})R_{3}(\Delta\psi)R_{1}(\epsilon_{A} + \Delta\epsilon), \\
	\label{PN_0} 
\end{equation}

where $\zeta_{A}, \theta_{A}, z_{A}$ are the Newcomb-Andoyer angles  (\cite{Lieske_1985}, \cite{iers10}), $\epsilon_{A}$ is the mean obliquity of the date, $\psi_{A}$ is the precession in longitude, $\Delta\psi$ and $\Delta\epsilon$ are corrections to $\psi_{A}$ , and  $\epsilon_{A}$ , correspondingly. 

Matrices $R_{1}$, $R_{2}$, and $R_{3}$ are defined as follows:

\begin{equation}
R_{1}(F) = 
\begin{pmatrix}
1 & 0 & 0 \\
0 & \cos F & \sin F\\
0 &  -\sin F  & \cos F 
\label{R1}
\end{pmatrix},  
\end{equation}

\begin{equation}
R_{2}(F) = 
\begin{pmatrix}
\cos F & 0 & -\sin F \\
0 & 1 & 0\\
\sin F &  0  & \cos F 
\label{R2}
\end{pmatrix},  
\end{equation}

\begin{equation}
R_{3}(F) = 
\begin{pmatrix}
\cos F & \sin F & 0 \\
-\sin F & \cos F & 0\\
0 &  0  & 1 
\label{R3}
\end{pmatrix}  .
\end{equation}

The following approximations are used for $P$ and $N$:

\begin{equation}
P \approx
\begin{pmatrix}
1 & (\zeta_{A} + z_{A}) & \theta_{A} \\
-(\zeta_{A} + z_{A}) & 1 & - \theta_{A}\zeta_{A}\\
  -\theta_{A} & -\theta_{A} z_{A}  & 1 
\label{P}
\end{pmatrix}  
\end{equation}

and

{\begin{equation}
N \approx
\begin{pmatrix}
1 & \Delta\psi\cos(\epsilon_{A}+\Delta\epsilon) & \Delta\psi\sin(\epsilon_{A} + \Delta\epsilon) \\
-\Delta\psi\cos\epsilon_{A} & \cos\Delta\epsilon & \Delta\epsilon\\
-\Delta\psi\sin\epsilon_{A} & -\Delta\epsilon  & \cos\Delta\epsilon
\label{N0}
\end{pmatrix}  
\end{equation}}

or

{\begin{equation}
N \approx
\begin{pmatrix}
1 & \Delta\psi\cos\epsilon_{A} & \Delta\psi\sin\epsilon_{A} \\
-\Delta\psi\cos\epsilon_{A} & 1 & \Delta\epsilon \\
-\Delta\psi\sin\epsilon_{A} & -\Delta\epsilon  & 1 
\label{N}
\end{pmatrix}.  
\end{equation}}

The combination of the six matrices in (\ref{PN_0}) may be presented in a simple form after all the approximations. 

\begin{strip}
    
\begin{equation}
PN \approx
\begin{pmatrix}
  1- (\zeta_{A} + z_{A})\Delta\psi\cos\epsilon_{A}
  - \Delta\psi\theta_{A} \sin\epsilon_{A} & \Delta\psi\cos\epsilon_{A}  +(\zeta_{A} + z_{A}) - \Delta\epsilon\theta_{A} & \Delta\psi\sin\epsilon_{A}  + \Delta\epsilon(\zeta_{A} + z_{A}) + \theta_{A}\\
  -\Delta\psi\cos\epsilon_{A}  -(\zeta_{A} + z_{A})  & 1 - (\zeta_{A} + z_{A})\Delta\psi\cos\epsilon_{A}  & \Delta\epsilon - \theta_{A}\zeta_{A} - (\zeta_{A} + z_{A}) \Delta\psi\sin\epsilon_{A} \\
  -\theta_{A} -\Delta\psi\sin\epsilon_{A} & -\Delta\epsilon - \theta_{A} z_{A} - \theta_{A}\Delta\psi\cos\epsilon_{A}  & 1 - \Delta\psi\theta_{A}\sin\epsilon_{A} -  \Delta\epsilon\theta_{A}z_{A} 
  
\label{PN0}
\end{pmatrix}  
\end{equation}

\end{strip}

For an alternative definition of the precession matrix $P$, we can write
\begin{equation}
  P = R_{1}({-\epsilon_{0}}) R_{3}(\psi_{A})R_{1}(\omega_{A})R_{3}(-\chi_{A}),\\
	\label{P2} 
\end{equation}

where $\epsilon_{0} = 84381''.406$ is the IAU2006 value for the obliquity at the epoch of J2000.0, $\omega_{A}$ is the precession in obliquity, $ $ $\chi_{A}$ is the precession of ecliptic along the mean equator of date \cite{Liu_2025}. Analytical expressions for the Newcomb-Andoyer angles are presented in \cite{Lieske_1985}, \cite{iers10}.

\subsection{New approach}

Another option to link GCRS and IRP is presented on Fig.~\ref{fig:concept}. Herewith the IRP is tied directly to the GCRS pole, and not intermediate pole is required. As a result, the transformation between the ICRS and IRP is similar to Eq.~(\ref{PN_0}):

\begin{equation}
  Q(t) = R_{2}(-X_{IRP})R_{1}(Y_{IRP})R_{3}(S), \\
	\label{B_1} 
\end{equation}

where $X_{IRP}$ and $Y_{IRP}$ are two angles that provide position of the instantaneous angular rotation vector with respect to some reference pole, and angle $S$, as mentioned below, is an analogue of the CIO locator from the IERS 2010 Conventions (\cite{iers10}), and should be referred as the IRP locator in this notation \citep{Liu_2026}. So, the full transformation matrix $A(t)$ in the new notation

\begin{figure}[ht]
    \centering
    \includegraphics[width=1.0\linewidth]{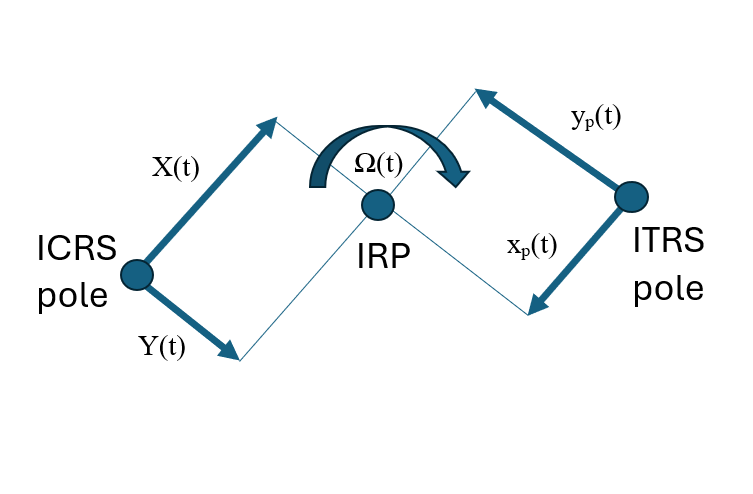}
 \caption{Representation of relevant poles of reference to Earth rotation for the proposed concept.}  
    \label{fig:concept}
\end{figure}

\begin{equation}
  A(t) =  R_{2}(-X_{IRP})R_{1}(Y_{IRP})R_{3}(S)R_{3}(-ERA)R_{2}(x_{p})R_{1}(y_{p}).\\
	\label{trans1} 
\end{equation}

Then, the matrix $Q_{IRP}(t)$

\begin{equation}
Q_{IRP}(t) =
\begin{pmatrix}
  1-\frac{X_{IRP}^2}{2}& -X_{IRP}Y_{IRP}  &  X_{IRP} \\
  0 & 1-\frac{Y_{IRP}^2}{2} & Y_{IRP}\\
  -X_{IRP} & -Y_{IRP} & 1-\frac{X_{IRP}^2+Y_{IRP}^2}{2}
	\label{B_2} 
\end{pmatrix}  R_{3}(S)
\end{equation}

or

\begin{equation}
Q_{IRP}(t) =
\begin{pmatrix}
  1-\frac{X_{IRP}^2+S^2}{2}& S-X_{IRP}Y_{IRP}  &  X_{IRP} \\
  -S  & 1-\frac{S^2+Y_{IRP}^2}{2} & Y_{IRP}\\
  -X_{IRP} + Y_{IRP}S & -Y_{IRP}-X_{IRP}S & 1-\frac{X_{IRP}^2+Y_{IRP}^2}{2}
	\label{B_3} 
\end{pmatrix}  
\end{equation}

So, the three angles could be linked to the Newcomb-Andoyer angles from Eq. (\ref{PN_0})

\begin{equation}
\begin{aligned}
  X_{IRP} & \approx  \theta_{A} + \Delta\psi\sin\epsilon_{A}  +
  (\zeta_{A} + z_{A})\Delta\epsilon  \\
 Y_{IRP} & \approx -\theta_{A}\zeta_{A} + \Delta\epsilon - (\zeta_{A} + z_{A})\Delta\psi\sin\epsilon_{A}  \\
 S & \approx (\zeta_{A} + z_{A}) + \Delta\psi\cos\epsilon_{A}
	\label{arguments} 
\end{aligned}
\end{equation}

As $\theta_{A} = \psi_{A}\sin\epsilon_{A} $  and $\zeta_{A}+z_{A} = \psi_{A}\cos\epsilon_{A}-\chi_{A}$, Eq. (\ref{arguments}) is converted to

\begin{equation}
\begin{aligned}
  X_{IRP} &  \approx  \psi_{A}\sin\epsilon_{A} + \Delta\psi\sin\epsilon_{A}  +
  \Delta\epsilon(\psi_{A}\cos\epsilon_{A}-\chi_{A})   \\
 Y_{IRP} & \approx  \Delta\epsilon -(\psi_{A}\cos\epsilon_{A}-\chi_{A})(\frac{1}{2}\psi_{A} \sin\epsilon_{A} +\Delta\psi\sin\epsilon_{A}) \\
 S & \approx (\psi_{A}\cos\epsilon_{A}-\chi_{A})  + \Delta\psi\cos\epsilon_{A}
	\label{arguments1} 
\end{aligned}
\end{equation}

where $p_{A} = \psi_{A}-\chi_{A}\cos\omega_{A}$ is known as the general precession in longitude (\cite{Lieske_1985}).

\hspace{1pt}

Angles
 $X_{IRP}$ and $Y_{IRP}$ are estimated directly from VLBI observations, so only angle $S$ should be addressed here. The precession part Eq.  (\ref{arguments1}) links $\theta_{A}$ and 
 $\zeta_{A}+z_{A}$. Therefore, one could approximates $S$ as follows

\begin{equation}
\begin{aligned}
 S & \approx \frac{X_{IRP} - \chi_{A}\tan\epsilon_{A} }{ \tan\epsilon_{A}+\Delta\epsilon}
	\label{S0} 
\end{aligned}
\end{equation}

As $\Delta\epsilon$ dominates in the expression for $Y_{IRP}$ this Eq~(\ref{S0}) may be replaced by its alteration consisting of the pair of angles $X_{IRP}$ and $Y_{IRP}$ and two values reflected the ecliptic orientation ($\chi_{A}$, $\epsilon_{A}$). 

\begin{equation}
\begin{aligned}
 S & \approx \frac{X_{IRP} - \chi_{A}\tan\epsilon_{A} }{ \tan\epsilon_{A} + Y_{IRP}}
	\label{S1} 
\end{aligned}
\end{equation}

 The new quantity $S$ (could be called as the IRP locator, by analogy with the CIO locator in the conventional approach) is used to rotate the reference system from the IRP origin to the GCRS origin along the IRP equator.

\hspace{1pt}

In IERS 2010 Conventions (\cite{iers10}) $R_{3}(\theta) = R_3(-ERA)$ accounts for the Earth Rotation Angle (ERA) around the CIP between the terrestrial intermediate origin and the CIO given by $\theta = \Omega \cdot UT1+\theta_{0}$ where is the $\Omega$ reference Earth rotation angular velocity, UT1 is the Earth rotation time, and $\theta_{0}$ a given constant angle, transforming Terrestrial Intermediate Reference System to Celestial Intermediate Reference System (CIRS).
In this manuscript angle $\theta$ plays the same role of the rotation angle, but now this is the rotation around IRP instead of CIP.

The system of equations (\ref{arguments}) is important because it links the traditional variables with new variables that could be estimated with the delay rate measured using geodetic VLBI.

One could present the non-diagonal elements of matrix (\ref{B_2}), also formulated with the traditional variables

\begin{equation}
\begin{aligned}
X_{IRP}S &\approx \theta_{A}(\zeta_{A} + z_{A}) + \theta_{A}\Delta\psi\cos\epsilon_{A}+ (\zeta_{A} + z_{A})\Delta\psi\sin\epsilon_{A}  \\
Y_{IRP}S &\approx \Delta\epsilon (\zeta_{A} + z_{A})  \\
X_{IRP}Y_{IRP} &\approx +\Delta\epsilon\theta_{A} \\
-X_{IRP}+Y_{IRP}S &\approx -\theta_{A} -\Delta\psi\sin\epsilon_{A} \\
Y_{IRP}+X_{IRP}S &\approx  \Delta\epsilon + \theta_{A}(\zeta_{A} + z_{A}) + \theta_{A}\Delta\psi\cos\epsilon_{A}
\end{aligned}
\end{equation}

\hspace{1pt}

To calculate $S$ consider Eq (\ref{B_2}) that could be split into asymmetric $Q_{IRP}'$ and symmetric $Q_{IRP}''$ parts

\begin{equation}
Q'_{IRP}(t) =
\begin{pmatrix}
  1-\frac{X_{IRP}^2}{2}& -\frac{X_{IRP}Y_{IRP}}{2}  &  X_{IRP} \\
  \frac{X_{IRP}Y_{IRP}}{2}  & 1-\frac{Y_{IRP}^2}{2} & Y_{IRP}\\
  -X_{IRP} & -Y_{IRP} & 1-\frac{X_{IRP}^2+Y_{IRP}^2}{2}
	\label{B_44} 
\end{pmatrix}  
\end{equation}

\begin{equation}
Q''_{IRP}(t) =
\begin{pmatrix}
  0& -\frac{X_{IRP}Y_{IRP}}{2}  &  0 \\
  -\frac{X_{IRP}Y_{IRP}}{2}  & 0 & 0\\
  0 & 0 & 0
	\label{B_5} 
\end{pmatrix}  
\end{equation}

with a small deformation elements $\Delta S$ which is contributing to the full transformation matrix $A(t)$

\begin{equation}
\begin{aligned}
  \Delta S  = -\frac{X_{IRP}Y_{IRP}}{2}   \\
	\label{stellar angle} 
\end{aligned}
\end{equation}

The matrix $Q''_{IRP}(t)$ (\ref{B_5}) is symmetrical, that is, by definition, it describes a tiny second-order deformation of the reference frame. Thus, the transition from IRP to GCRS within the framework of the old concept, in addition to three rotation angles, also contains a small deformation.

\end{appendix}

\end{document}